\documentclass{aa}

\usepackage{graphicx}
\usepackage{mathabx}
\usepackage{txfonts}
\begin{document}

   \title{Recycling of the first atmospheres of embedded planets: \newline Dependence on core mass and optical depth}
   \titlerunning{Recycling of embedded planets’ first atmospheres}

   \author{T.~W.~Moldenhauer
          \inst{1}
          \and
          R.~Kuiper
          \inst{2,1}
          \and
          W.~Kley
          \inst{1}
          \fnmsep
          \thanks{deceased}
          \and
          C.~W.~Ormel
          \inst{3}
         }

   \institute{Institut für Astronomie und Astrophysik, Universität Tübingen,
              Auf der Morgenstelle 10, 72076 Tübingen, Germany
              \and
              Zentrum für Astronomie der Universität Heidelberg, Institut für Theoretische Astrophysik, Albert-Ueberle-Straße 2, 69120 Heidelberg, Germany
              \and
              Department of Astronomy, Tsinghua University,
              30 Shuangqing Rd, 100080 Haidian District, Beijing, China\\\\
              \email{tobias.moldenhauer@uni-tuebingen.de}
             }
   \date{Received August 1, 2021; accepted February  7, 2022}

 
  \abstract
   {
   Recent observations found close-in planets with significant atmospheres of hydrogen and helium in great abundance. These are the so-called super-Earths and mini-Neptunes.
   Their atmospheric composition suggests that they formed early during the gas-rich phase of the circumstellar disk and were able to avoid becoming hot Jupiters.
   As a possible explanation, recent studies explored the recycling hypothesis and showed that atmosphere-disk recycling is able to fully compensate for radiative cooling and thereby halt Kelvin-Helmholtz contraction to prevent runaway gas accretion.
   }
   {
   To understand the parameters that determine the efficiency of atmospheric recycling, we extend our earlier studies by exploring the effects of the core mass, the effect of circumstellar gas on sub-Keplerian orbits (headwind), and the optical depth of the surrounding gas on the recycling timescale.
   Additionally, we analyze their effects on the size and mass of the forming atmosphere.
   }
   {
   We used three-dimensional (3D) radiation-hydrodynamic simulations to model a local shearing box centered on the planet.
   Our planet is located at a separation of $a_p = 0.1 \, \mathrm{au}$ from its solar-type host star, and we scanned the core mass range from $1$ to $10 \, M_\mathrm{Earth}$.
   In order to measure and track the recycling of the atmosphere, we employed tracer particles as well as tracer fluids after thermodynamic equilibrium was reached.
   }
   {
   For the explored parameter space, all simulations eventually reach an equilibrium where heating due to hydrodynamic recycling fully compensates radiative cooling.
   In this equilibrium, the atmosphere-to-core mass ratio stays well below $10 \, \%$, preventing the atmosphere from becoming self-gravitating and entering runaway gas accretion.
   Higher core masses cause the atmosphere to become turbulent, which further enhances recycling.
   Compared to the core mass, the effect of the headwind on the recycling timescale is negligible.
   The opacity has no significant effect on the recycling timescale, which demonstrates that the Kelvin-Helmholtz contraction timescale and the atmosphere-disk recycling timescale are independent of each other.
   }
   {
   Even for our highest core mass of $10 \, M_\mathrm{Earth}$, atmosphere-disk recycling is efficient enough to fully compensate for radiative cooling and prevent the atmosphere from becoming self-gravitating.
   Hence, in-situ formation of hot Jupiters is very unlikely, and migration of gas giants is a key process required to explain their existence.
   Our findings imply that atmosphere-disk recycling is the most natural explanation for the prevalence of close-in super-Earths and mini-Neptunes.
   }


\maketitle

\section{Introduction}

Recent observations by the \textit{Kepler} mission and the Transiting Exoplanet Survey Satellite (\textit{TESS}) found close-in super-Earths and mini-Neptunes in great abundance \citep{Weiss_2018}.
One out of three solar-type stars may host such a planet \citep{Winn_2015}.
These planets are small, several Earth radii in size, orbit close to their host stars, with a typical or mean distance of 0.1~au, and have an atmosphere that mainly consists of hydrogen and helium, which points to an early formation \citep{Wu_2013}.
Because of the short dynamical timescale and higher surface density at such close orbits \citep{Lee_2014}, the cores should have formed early enough for these protoplanets to have had enough time to accrete massive enough atmospheres for them to become self-gravitating and enter runaway accretion to become hot Jupiters \citep{Batygin_2016}.
In order to explain the high abundance of super-Earths and mini-Neptunes, an efficient mechanism to prevent runaway gas accretion is necessary.
Multiple solutions have been suggested, including a high-opacity atmosphere \citep{Lee_2014}, late formation \citep{Lee_2015}, atmosphere-disk recycling \citep{Ormel_2015b, Cimerman_2017, Bethune_2019b, Ali-Dib_2020, Moldenhauer_2021}, and gap opening \citep{Fung_2018, Ginzburg_2019}.

In the core-accretion model, giant planets form by accreting gas from the circumstellar disk onto their planetary cores \citep{Perri_1974, Mizuno_1978, Pollack_1996}.
Computationally, the thermal evolution of the atmosphere of these protoplanets is most conveniently calculated in one radial dimension (1D).
A quasi-steady state is reached when a luminosity source is present at the inner boundary (the interface between core and envelope), which is traditionally attributed to the accretion of solids.
In this approximation, the protoplanetary atmosphere\footnote{We use the words envelope and atmosphere interchangeably.} becomes a function of mass and solid accretion rate, but is independent of time \citep{Rafikov_2006}.
Nevertheless, more massive protoplanetary cores accrete more nebular gas.
Loosely speaking, the point at which gas accretion overtakes solid accretion, the so-called crossover point, marks the onset of runaway gas accretion \citep{Bodenheimer_1986, Pollack_1996, Papaloizou_1999}.
Previous 1D models calculated a critical core mass of $M_\mathrm{crit} \sim 5 {-} 20 \, M_\mathrm{Earth}$ as a requirement for the occurrence of runaway gas accretion \citep{Mizuno_1980, Ikoma_2000}.

The situation changes when solid accretion terminates and the core mass is fixed.
The atmosphere then provides its own luminosity by Kelvin-Helmholtz contraction.
For close-in super-Earths and mini-Neptunes, this situation can be expected because any local source of solids (planetesimals) will be accreted on timescales much shorter than the disk lifetime.
In addition, for higher-mass protoplanetary cores ($M_c \sim 20 \, M_\mathrm{Earth}$), pebble accretion may also be halted once the pebble isolation mass is reached \citep{Lambrechts_2014,Ataiee_2018,Bitsch_2018}.
Detailed 1D studies have consistently found that the Kelvin-Helmholtz timescale is a strong function of the opacity in the outer radiative zone \citep{Ikoma_2000,Lee_2014,Hori_2010,Chachan_2021}.
It has been typical to assume interstellar medium (ISM)-like dust-to-gas ratios and grain properties, in which case the opacity is overwhelmingly determined by the dust, particularly in the outer disk.
However, grains are expected to aggregate and settle, giving rise to the conclusion that envelopes should be modeled as grain-free (\citet{Mordasini_2014,Ormel_2014}).
In this situation, the timescale on wich runaway gas accretion is to be achieved drops  below 1\,Myr for core masses that exceed 2\,$M_\oplus$ \citep{Hori_2010}.

Despite these concerns, 1D models are still widely employed to explain the preponderance of super-Earths and mini-Neptunes \citep{Ida_2004, Mordasini_2009, Schlecker_2020}, simply assuming a sufficiently high opacity and a long period in which solids are accreted, or by invoking planetary migration.
These 1D models have already demonstrated that the opacity and heating through accretion of solids are important parameters for the evolution of protoplanetary atmospheres.
However, a comparison of ALMA observations to current gas accretion models suggested that gas accretion needs to be suppressed by at least one order of magnitude to match the observed planet population \citep{Nayakshin_2019}.
Some studies have argued that a supersolar metallicity of the gas \citep{Lee_2014} or convective overshooting \citep{Ali-Dib_2020, Johansen_2020} might slow down or halt the Kelvin-Helmholtz contraction.
A more straightforward answer may come from hydrodynamics, however.
In the past decade, a series of hydrodynamical simulations have indicated that the 1D assumption of an atmosphere isolated from its surroundings is no longer applicable in 3D \citep{Ormel_2015b,Fung_2015,Cimerman_2017,Kurokawa_2018,Kuwahara_2020}.

However, for geometric reasons, 1D (and 2D) models fail to capture the 3D flow structure, which enables recycling of low-entropy atmospheric gas with high-entropy circumstellar gas.
This makes capturing recycling effects computationally expensive, which generally means that these effects are ignored.
In the 3D picture, circumstellar gas is accreted through the vertical direction and can be removed in the midplane \citep{Machida_2008,Tanigawa_2012,Szulagyi_2016},
and 3D studies found that there is no clear boundary between the atmospheric and circumstellar gas flow and gas is constantly exchanged between both regions, it is recycled.
High-entropy circumstellar gas replaces low-entropy atmospheric gas and thereby heats the cooling atmosphere, providing an effective heat source that counteracts radiative cooling \citep{Ormel_2015b, Cimerman_2017}.
In recent work, we have shown that this hydrodynamic heating effect allows reaching a true steady state \citep{Moldenhauer_2021}.
Consequently, for close-in planetary cores, there is no atmosphere in the traditional sense of gravitationally bound gas.
Therefore, we introduce another definition for the planetary atmosphere that is based on the recycling time, which is the time it takes gas to flow back from the vicinity of the core to the circumstellar disk.
Gas that has a significantly longer recycling time than its surroundings, that is, the gas that orbits the core instead of the star for several orbital periods until it is replenished, is defined as the atmosphere (see section \ref{sec:length_scales} for details).

To test the efficacy of the recycling hypothesis, we employ a gas opacity in our simulations that is several orders of magnitude lower than a realistic opacity \citep{Ferguson_2005}.
By choosing an artificially low opacity, we focus our analysis on the recycling mechanism.
A 1D simulation with the same parameters would reach $M_\mathrm{atm} \sim M_c$ in fewer than 10,000 orbits \citep{Moldenhauer_2021}.
Additionally, a convenient side effect of the shorter cooling timescale is that it saves computational resources by enabling the simulations to reach equilibrium faster.
By deliberately using a very low opacity (and no accretion luminosity), \cite{Moldenhauer_2021} showed that recycling on its own is capable of fully compensating for radiative cooling and keeping the atmosphere of an Earth-mass planet at $a_p = 0.1 \, \mathrm{au}$ in thermodynamic equilibrium.
In this paper, we continue this work by investigation the effects of the protoplanetary core mass on the recycling timescale.

In addition to the core mass, another important parameter is the so-called headwind.
The disk headwind is defined as the amount by which the gas rotational velocity deviates from Keplerian.
Because of the radial pressure gradient in the circumstellar disk, the gas is partially pressure supported.
This pressure support acts against the stellar gravity, causing the gas to rotate around the star at sub-Keplerian speeds.
As the gas rotates at a lower velocity than the planet, the planet effectively experiences a headwind.
The headwind strongly affects the boundary region between the planetary atmosphere and the circumstellar disk, potentially altering the recycling process. 
A stronger headwind leads to a more asymmetric flow pattern and an overall smaller atmosphere \citep{Ormel_2013}.

We continue from \cite{Moldenhauer_2021} and use local radiation-hydrodynamic (RHD) simulations to perform a parameter study and investigate the influence of the core mass, the opacity, and the headwind on the recycling process.
Our study focuses on close-in super-Earths and mini-Neptunes.
Simulations at larger separations require significantly more computational resources because the dynamical timescale is longer than the computational time step, which is limited by the inner boundary.
All our simulations use the surface of the solid protoplanetary core as its inner boundary, which is independent of the separation.
We reserve this parameter for a future study.
Here, we show that for the explored parameter range, all simulations eventually reach thermodynamic equilibrium where radiative cooling is fully compensated for by hydrodynamic recycling.
In this thermodynamic equilibrium, atmospheric growth halts completely.
Additionally, we study the effects that the analyzed parameters have on the recycling timescale, and we study how the structure and mass of the resulting atmosphere is changed.

\begin{table*}
   \centering
   \begin{tabular}{c c c c c c c c c c c}
      \hline\hline
      Name & $M_\mathrm{c}$ / $M_\Earth$ & $\mathcal{M}_\mathrm{hw}$ & $\kappa$ / $\mathrm{cm}^2 \, \mathrm{g}^{-1}$ & $t_\mathrm{inj} / \Omega_K^{-1}$ & $M_\mathrm{H} / M_\mathrm{c}$  & $M_\mathrm{B} / M_\mathrm{c}$ &  $M_\mathrm{atm} / M_\mathrm{c}$ & $R_\mathrm{H} / R_\Earth$ & $R_\mathrm{B} / R_\Earth$ & $R_\mathrm{atm} / R_\Earth$\\
      \hline
      M1 & 1 & 0 & $10^{-4}$ & $4$ & 2.70 \% & 1.37 \% & 0.20 \% & 23.5 & 17.3 & 6.0\\
      M2 & 2 & 0 & $10^{-4}$ & $4$ & 3.52 \% & 5.10 \% & 0.42 \% & 29.6 & 35.4 & 8.2\\
      M5 & 5 & 0 & $10^{-4}$ & $4$ & 5.05 \% & 21.38 \% & 1.32 \% & 40.1 & 98.9 & 14.7\\
      M10 & 10 & 0 & $10^{-4}$ & $4$ & 4.43 \% & 40.85 \% & 1.91 \% & 50.6 & 212.7 & 24.6\\
      \hline
      M1-HW & 1 & 0.1 & $10^{-4}$ & $4$ & 2.71 \% & 1.37 \% & 0.14 \% & 23.5 & 17.3 & 4.8\\
      M2-HW & 2 & 0.1 & $10^{-4}$ & $4$ & 3.52 \% & 5.10 \% & 0.37 \% & 29.6 & 35.4 & 7.6\\
      M5-HW & 5 & 0.1 & $10^{-4}$ & $4$ & 5.02 \% & 21.21 \% & 1.04 \% & 40.1 & 98.9 & 11.7\\
      M10-HW & 10 & 0.1 & $10^{-4}$ & $4$ & 4.37 \% & 40.71 \% & 1.87 \% & 50.6 & 212.7 & 24.6\\
      \hline
      M1-HO & 1 & 0 & $10^{-3}$ & $4$ & 2.68 \% & 1.35 \% & 0.15 \% & 23.5 & 17.3 & 5.3\\
      M1-LO & 1 & 0 & $10^{-5}$ & $4$ & 2.90 \% & 1.40 \% & 0.44 \% & 23.5 & 16.5 & 8.1\\
      \hline
      M1-SLOW & 1 & 0 & $10^{-4}$ & $400$ & 2.70 \% & 1.37 \% & 0.20 \% & 23.5 & 17.3 & 6.0
   \end{tabular}
   \caption{Simulations and their respective atmospheric mass inside different radii. Bondi and Hill are abbreviated with B and H, respectively. $R_\mathrm{atm}$ is the radius at which the azimuthal velocity is the strongest component in the midplane, i.e., at which the flow of the circumstellar disk starts and the atmospheric flow ends (see section \ref{sec:length_scales}). The simulations are labeled MX, where X is the mass of the core in units of $M_\mathrm{earth}$. The suffixes describe alterations compared to the standard setup: HW means that a headwind is included, HO and LO stand for a higher and lower opacity, respectively. Finally, M1-SLOW is a simulation with an injection time that is 100 times longer.}
   \label{table:simulations}
\end{table*}

\section{Model}

For our simulations, we built upon the code and parameters used by \cite{Cimerman_2017} and \cite{Moldenhauer_2021}.
We used RHD on a 3D spherical grid in a local shearing box approach, centered on the planet.
Figure \ref{fig:sketch} outlines the structure of the setup.
The separation of the planet from its host star is $a_p=0.1 \, \mathrm{au,}$ and we adopt a corresponding ambient midplane density of $\rho_0 = 6\cdot10^{-6} \, \mathrm{g} \, \mathrm{cm}^{-3}$ and temperature of $T_0 = 1000 \, \mathrm{K}$ following \cite{Lee_2014}, in accordance with the minimum-mass extrasolar nebula \citep[MMEN;][]{Chiang_2013}.
In the optically thick regime, \cite{Cimerman_2017} showed that the final mass of the atmosphere is rather independent of opacity or total optical depth.
For the parameter scan, we therefore used a single value of opacity, $\kappa = 10^{-4} \, \mathrm{cm}^2 \, \mathrm{g}^{-1}$, in order to keep the cooling timescale short and thus save on computation time;
however, in section \ref{sec:result_opacity} we compare different opacities to explore the effect of optical depth.
The simulation domain covers the upper hemisphere, and we did not use any smoothing for the gravitational potential of the planet as we resolve the rocky planetary surface.
To explore the effect of the core mass on the flow structure as well as the atmospheric mass, we performed simulations with core masses $M_c \in [1,2,5,10] \, M_\mathrm{Earth}$.
For the rocky cores, we assumed a mean density of $\rho_c = 5 \, \mathrm{g} \, \mathrm{cm}^{-3}$.
Models are labeled MX, where X is the core mass in units of Earth masses.
The M1 simulation is identical to the setup described in \cite{Moldenhauer_2021}.
All simulations are listed in Table \ref{table:simulations}.

\subsection{Headwind}

Because of the radial thermal pressure gradient through the circumstellar disk, the gas experiences an outward force, $\boldsymbol{F}_p = - \nabla p$, that acts against the stellar gravity.
This results in a sub-Keplerian rotation of the gas.
For solid bodies such as planets, on the other hand, this force is negligible.
The planet therefore moves faster than the surrounding gas and thus experiences a headwind along its orbital direction.
As we did not model the full circumstellar disk, we artificially added this headwind to our shearing box setup.
In a global simulation, this effect would be naturally included.
We parameterized the headwind in our simulations in terms of the Mach number, $\mathcal{M}_\mathrm{hw}$, where
\begin{align}
   v_\mathrm{hw} &= \mathcal{M}_\mathrm{hw} \cdot c_\mathrm{iso}
   \label{eq:headwind}
\end{align}
is the gas velocity of the headwind and $c_\mathrm{iso}$ is the isothermal sound speed for the initial and boundary temperature, $T_0$.
From the temperature profile we used for our simulations, we can calculate the headwind Mach number as
\begin{align}
    \mathcal{M}_\mathrm{hw} &= 0.033 \left( \frac{a}{0.1 \mathrm{au}} \right)^{2/7}.
\end{align}
For the minimum-mass solar nebula (MMSN), we obtain a slightly lower headwind Mach number \citep{Kuwahara_2020},
\begin{align}
    \mathcal{M}_\mathrm{hw} &= 0.028 \left( \frac{a}{0.1 \mathrm{au}} \right)^{1/4}.
\end{align}
However, to clearly assess the effects of the headwind and show its impact clearly, we followed \cite{Ormel_2015b} and chose a headwind Mach number of $\mathcal{M}_\mathrm{hw} = 0.1$.
Models that include a headwind are marked by the abbreviation HW in their names.

\subsection{Length scales}
\label{sec:length_scales}

There are two important length scales in protoplanetary atmospheres.
The first is the Hill radius,
\begin{align}
   R_\mathrm{Hill} &:= a \sqrt[3]{\frac{M_\mathrm{c}}{3M_\star}},
\end{align}
which is the radius of the region in which the planetary gravity dominates the stellar gravity.
The second is the Bondi radius,
\begin{align}
   R_\mathrm{Bondi} &:= \frac{2G M_\mathrm{c}}{c_s^2},
\end{align}
which is the radius at which the escape velocity equals the speed of sound.
To calculate the Bondi radius, we used the spherically averaged local sound speed.
The Bondi radius is therefore time dependent and a local length scale.
The Hill radius, on the other hand, is constant in time and a global length scale.
We emphasize that in the 3D setup, physical relevant physical quantities as well as the flow field are continuous across $R_\mathrm{Hill}$ and $R_\mathrm{Bondi}$, but we nevertheless included them for comparison purposes.

For an atmosphere that fully recycles, the traditional definition of an atmosphere as the region in which the gas is gravitationally bound is no longer meaningful.
Therefore, in addition to the Hill and Bondi radii, we defined a third radius $R_\mathrm{atm}$ to analyze the size and mass of the protoplanetary atmosphere,
\begin{align}
    R_\mathrm{atm} &:= \operatorname{max} \left\{ r \, \middle| \, v_\varphi |_{\theta = \pi/2} > \operatorname{max}(v_r, v_\theta) \right\}.
\end{align}
$R_\mathrm{atm}$ is the maximum radius at which the azimuthal velocity component is the largest component of the velocity field in the midplane, that is, where the gas rotates around the core instead of the star.
This radius is a good measurement for the size of the atmosphere as it indicates where the circular atmospheric flow transitions into the circumstellar shearing flow.
In the following analyses, the 2D cut figures include this radius as a magenta circle.
The resulting sphere matches the size of the atmosphere well that is obtained from the regions that are recycled on longer timescales, $t_\mathrm{rec} >100 \, \Omega_K^{-1}$.
Throughout this work, "atmosphere" refers to the region inside $R_\mathrm{atm}$.
At the time of disk dispersal, this region might form a proper bound atmosphere.
However, because our simulations do not cover the dispersal of the disk, this remains work for a further investigation.

\begin{figure*}[ht]
   \centering
   \includegraphics[width=0.98\hsize]{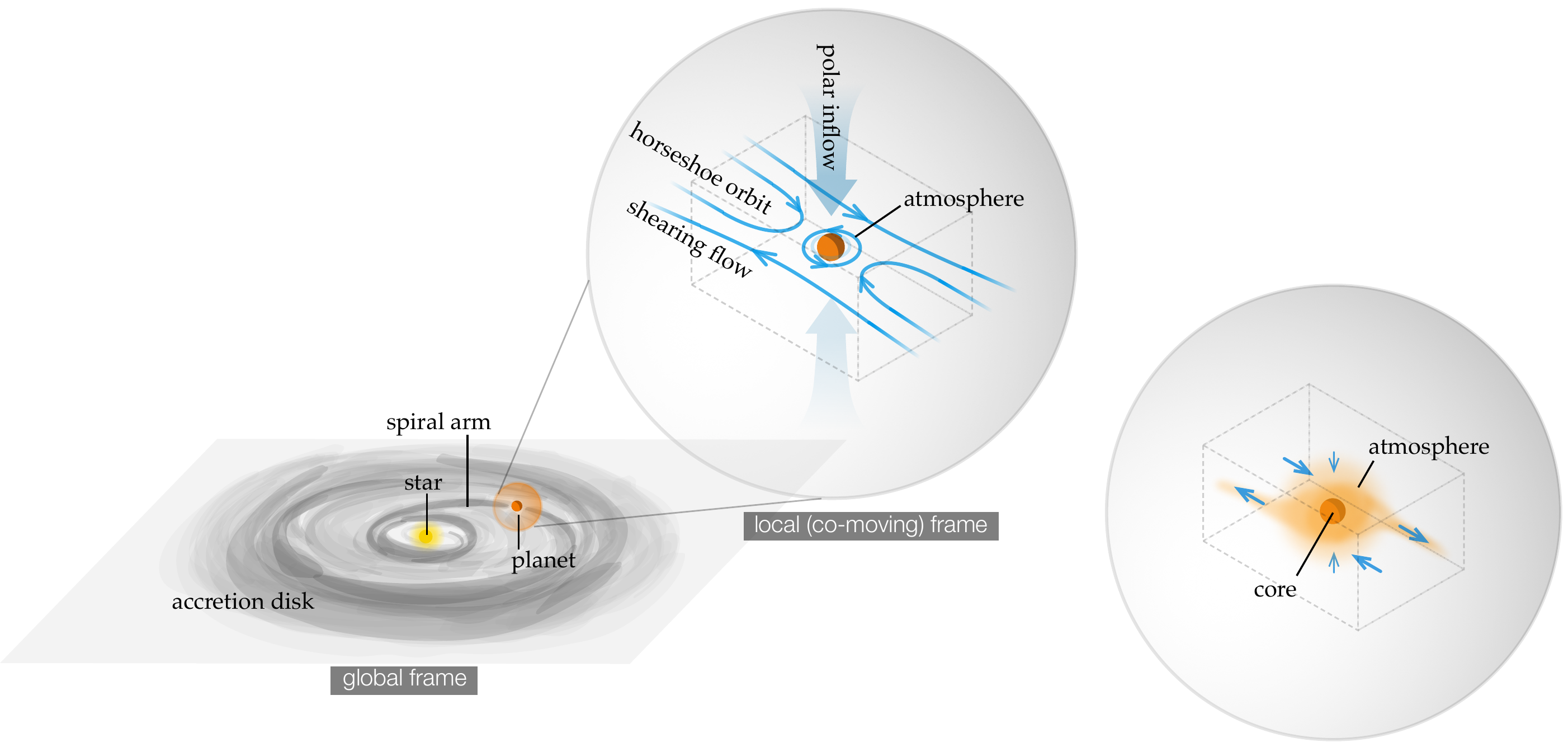}
   \caption{
   Outline of the setup.
   We introduce a rocky core and simulate the protoplanetary atmosphere in a local frame, while we keep the conditions of the circumstellar disk constant in time.
   The planetary core accumulates gas to form an atmosphere.
   In addition to polar inflow, the two horseshoe orbits that form along the orbital direction of the planet constantly exchange high-entropy circumstellar with low-entropy atmospheric gas.
   This process results in thermodynamic equilibrium in which radiative cooling is fully compensated for by atmosphere-disk recycling.
   We thank André Oliva for drawing this sketch.
   }
   \label{fig:sketch}%
\end{figure*}

\section{Numerics}

\subsection{Radiation hydrodynamics}

We used the hydrodynamics module of the open-source code PLUTO \citep{Mignone_2007}, version 4.1, and the flux-limited diffusion solver described in \cite{Kuiper_2009} and \cite{Kuiper_2020}.
PLUTO solves conservation equations: conservation of mass, momentum, and total energy,
\begin{align}
    \frac{\partial \rho}{\partial t} + \nabla \cdot (\rho \boldsymbol{v}) &= 0,\\
    \label{eq:momentum}
    \frac{\partial \rho \boldsymbol{v}}{\partial t} + \nabla ( \rho \boldsymbol{v} \cdot \boldsymbol{v} + p) - \rho \boldsymbol{a} &= 0,\\
    \label{eq:energy}
    \frac{\partial E}{\partial t} + \nabla \cdot [(E+p) \cdot \boldsymbol{v}] + \nabla \cdot \boldsymbol{F} - \rho \boldsymbol{v} \cdot \boldsymbol{a} &= 0.
\end{align}
Here, $\rho$ is the gas density, $p$ is the gas pressure, $\boldsymbol{v}$ is the gas velocity, and $E=E_\mathrm{th} + E_\mathrm{kin}$ is the total energy of the gas.
$\boldsymbol{a}$ represents external accelerations, and $\boldsymbol{F}$ is the radiation flux from the flux-limited diffusion solver.

Because our simulations use the advantages of a shearing box around the planet, which is not an inertial frame of reference, we have to include forces of inertia that act on the gas.
First, we include the Coriolis acceleration,
\begin{align}
   \boldsymbol{a}_\mathrm{cor} &= -2 \boldsymbol{\Omega}_K \times \boldsymbol{v},
   \label{eq:coriolis}
\end{align}
where $\boldsymbol{\Omega}_K = \Omega_K \cdot \boldsymbol{e}_z$ is the Keplerian frequency at the location of the planet pointing in the z-direction.
The tidal acceleration is the sum of the stellar gravity, $-GM_\star / a^2 $, and the centrifugal acceleration, $\Omega_K^2 a$.
Here $a$ is the distance of the gas to the star.
Using the Keplerian frequency, $\Omega_K = GM_\star / a_p^3$, at the orbital position of the planet, $a_p$, the linearized approximation of the tidal acceleration calculates as
\begin{align}
   \boldsymbol{a}_\mathrm{tid} &\approx 3\Omega_K^2 \Delta x \cdot \boldsymbol{\hat e}_x,
\end{align}
where $\Delta x = a - a_p$ is the relative orbital position of the gas with respect to the planet.
Finally, for the simulations that include a headwind, we add an additional acceleration,
\begin{align}
   \boldsymbol{a}_\mathrm{hw} &= 2 \mathcal{M}_\mathrm{hw} c_\mathrm{iso} \Omega_K \cdot \boldsymbol{\hat e}_x,
   \label{eq:headwind_acc}
\end{align}
to account for the net effect of the radial thermal pressure gradient.
Together with the gravitational acceleration by the planetary core, these accelerations add up to the total external acceleration, $\boldsymbol{a}$, in equation \ref{eq:momentum} and \ref{eq:energy}.

In the radial direction, the computational domain extends from the planetary core $r_c$ to $r_\mathrm{max} = 5H = 5 h \cdot a_p$, where $h = c_\mathrm{iso} / v_K = 0.02$ denotes the aspect ratio and $H$ is the pressure scale height of the circumstellar disk at the location of the planet.
In the polar and azimuthal directions, the domain covers the hemisphere above the midplane, and symmetry at the midplane is implied by the boundary conditions.
For all simulations, we used a resolution of 128 logarithmically spaced cells in the radial, 32 cells in the polar, and 128 in the azimuthal direction.

Simulations were initialized with the circumstellar disk shear flow and a planar uniform but vertically stratified density,
\begin{align}
   \rho_{t=0}(\vec r) = \rho_0 \cdot \exp\left(- \frac{1}{2} \frac{z^2}{H^2} \right).
\end{align}
The radially inner boundary at the planetary core is reflective, and the values at the outer boundary were kept constant at their initial values.
For the boundary at the vertical axis, $\theta=0$, we used the so-called $\pi$-boundary \citep{Ormel_2015b}.
Here, the ghost cells, which have an effective negative $\theta$ value, are filled with quantities from $\varphi + \pi$,
\begin{align}
   A(-\theta, \varphi) = A(+\theta, \varphi+\pi).
   \label{eq:pi_boundary}
\end{align}
Even though the interface of this boundary actually has an areal extent of zero, the values of the ghost cells have to be specified in the PLUTO code and are used in the reconstruction step of the hydrodynamics solver to compute the slope of grid-cell-centered variables.

To prevent the occurrence of initial shocks at the planetary core surface, we multiplied the planetary gravity and the headwind magnitude with a switch-on factor,
\begin{align}
   \alpha_\mathrm{inj}(t) &= 1-\exp\left( -\frac{1}{2} \frac{t^2}{t_\mathrm{inj}^2} \right),
   \label{eq:injection_factor}
\end{align}
where $t_\mathrm{inj} = 4 \, \Omega_K^{-1}$ is the injection time.
To confirm that the chosen injection time $t_\mathrm{inj} = 4 \, \Omega_K^{-1}$ does not affect the equilibrium solution of our simulation, we ran an additional simulation M1-SLOW, whose parameters are identical to those of the M1 simulation, but the injection time is much longer, $t_\mathrm{inj} = 400 \, \Omega_K^{-1}$.
With this long injection time, the density and thus the temperature of the atmosphere build up more slowly.
This might give the atmosphere more time to cool before recycling starts, and it might result in a different equilibrium of cooling and recycling.
However, this is not the case, as Table \ref{table:simulations} shows.
The mass and thermodynamics of the final atmosphere are independent of the chosen injection time.
Finally, for simulations including a headwind, we added the headwind velocity, $v_\mathrm{hw}$, to the shearing velocity at the radially outer boundary.

\begin{figure*}[ht]
   \centering
   \includegraphics[width=0.98\hsize]{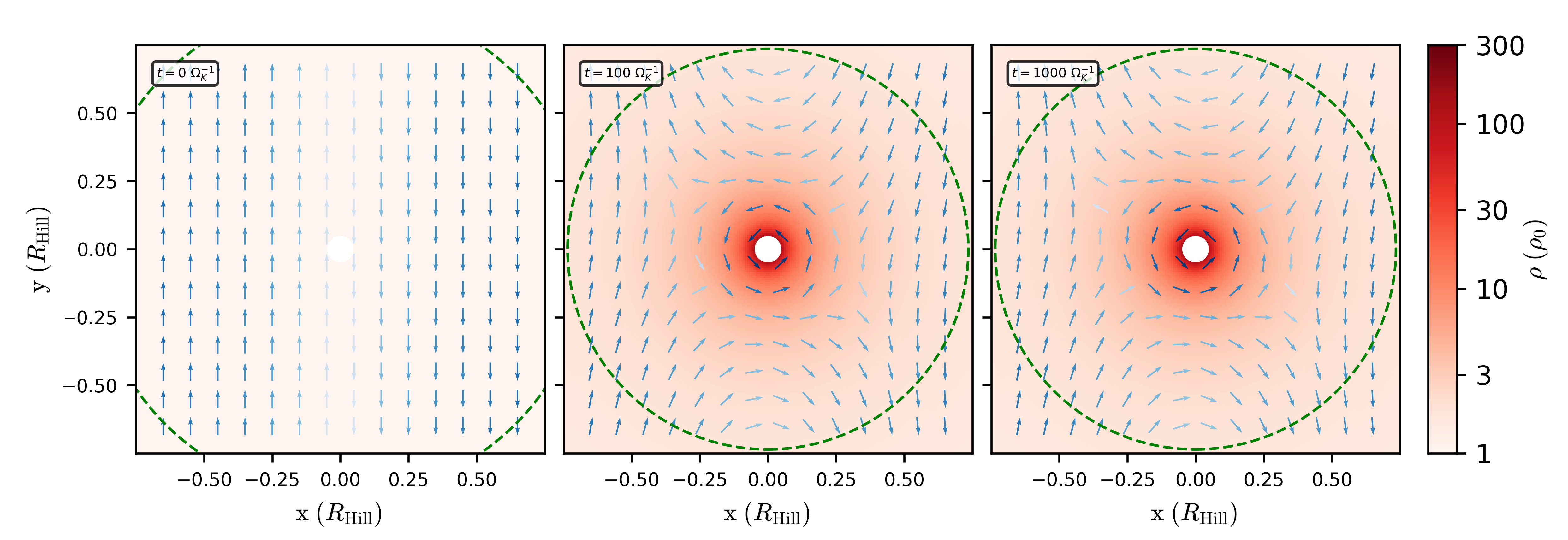}
   \caption{
      Development of the flow structure in the midplane for the M1 simulation.
      The computation domain is initialized with a uniform density and the shear flow of the circumstellar disk.
      The planetary core accretes gas, and a protoplanetary atmosphere forms around the core.
      In the orbital direction of the planet, horseshoe orbits form, centered on the sign change of the shear flow.
      Eventually, the simulation reaches thermodynamic equilibrium, where radiative cooling is fully compensated for by atmosphere-disk recycling.
      In this equilibrium the gas temperature and the gas density become constant in time.
      The dashed green circle is the Bondi radius calculated from the local temperature.
      The Bondi radius shrinks with time because the atmospheric gas heats up when it is compressed.
      After $\sim 100 \, \Omega_K^{-1}$ the simulation reaches thermodynamic equilibrium.
   }
   \label{fig:M1_density_evolution}%
\end{figure*}

\begin{figure*}[ht]
   \centering
   \includegraphics[width=0.49\hsize]{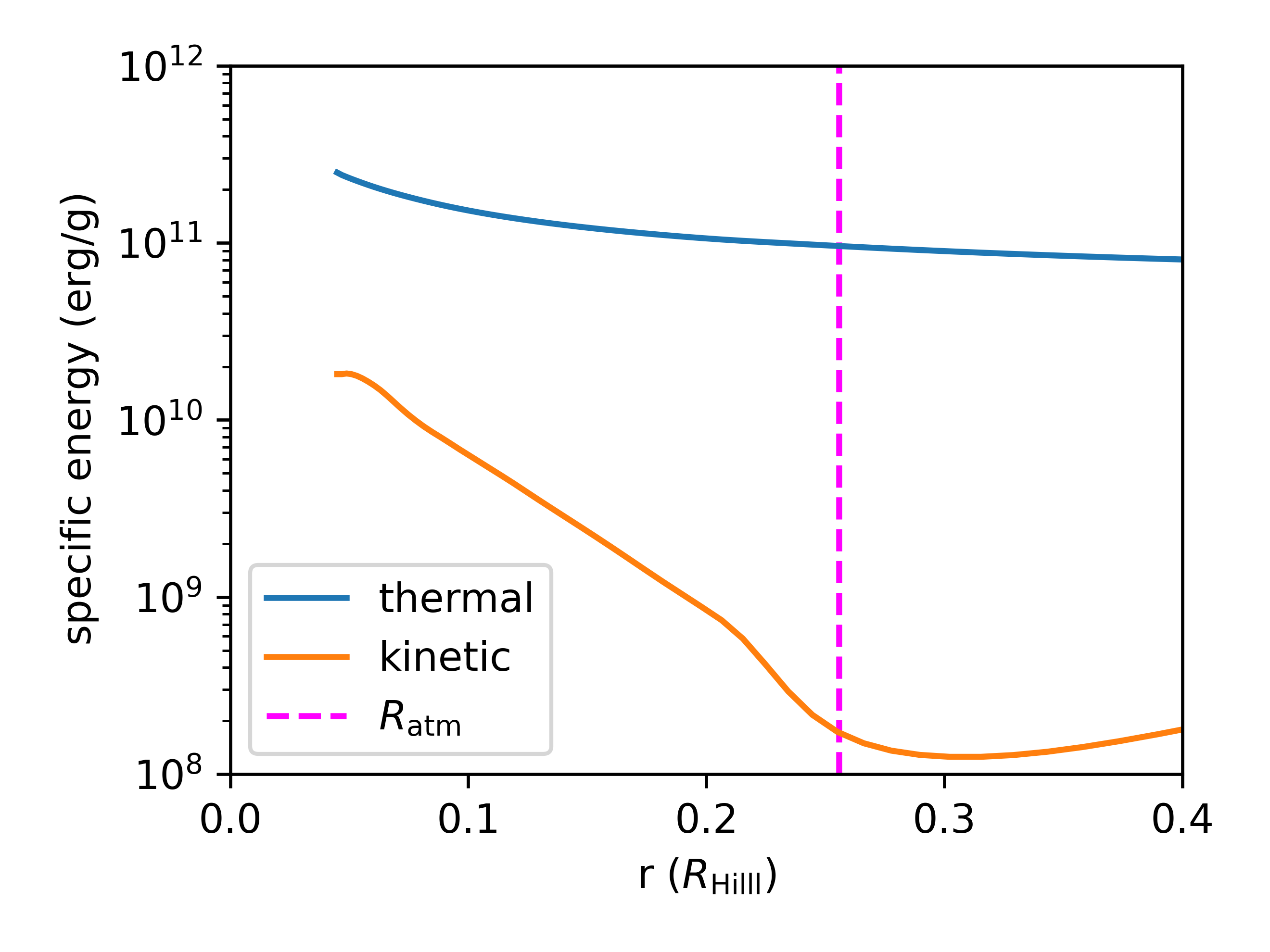}
   \includegraphics[width=0.49\hsize]{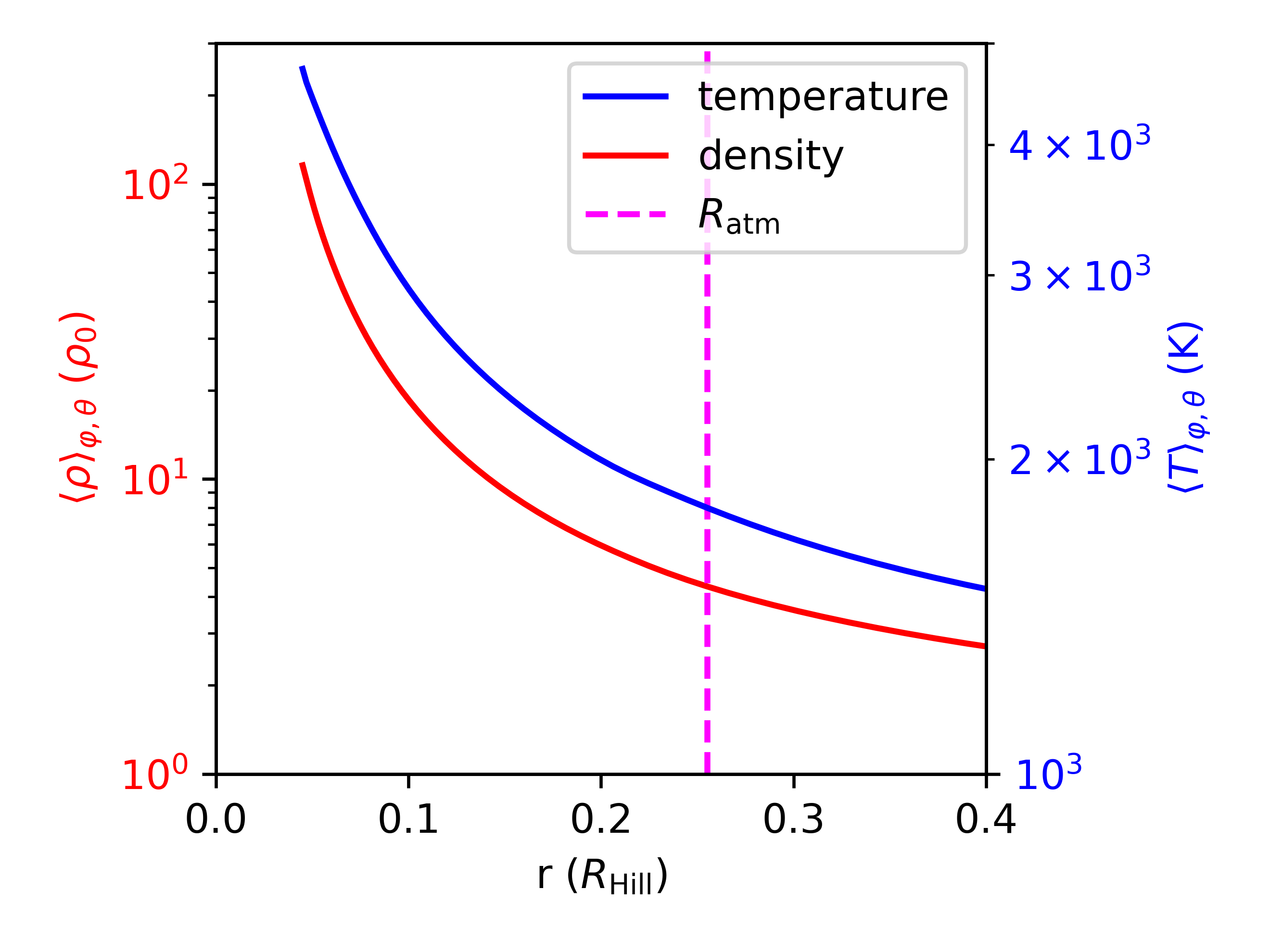}
   \caption{
   Radial 1D-averaged profiles for M1 simulation.
   Left: Thermal and kinetic specific energy profiles.
   Right: Density and temperature profiles.
   The dashed magenta vertical line is $R_\mathrm{atm}$ as defined in Sect. \ref{sec:length_scales}. The thermal energy is more than one order of magnitude higher than the kinetic energy, i.e.,~the atmosphere is pressure supported. Inside $R_\mathrm{atm}$ , the kinetic energy is significantly higher than outside. This is because of the azimuthal velocity component: Inside $R_\mathrm{atm}$, the gas rotates around the core.
   }
   \label{fig:M1_energy_density_temperature}%
\end{figure*}

\begin{figure*}[ht]
   \centering
   \includegraphics[width=0.49\hsize]{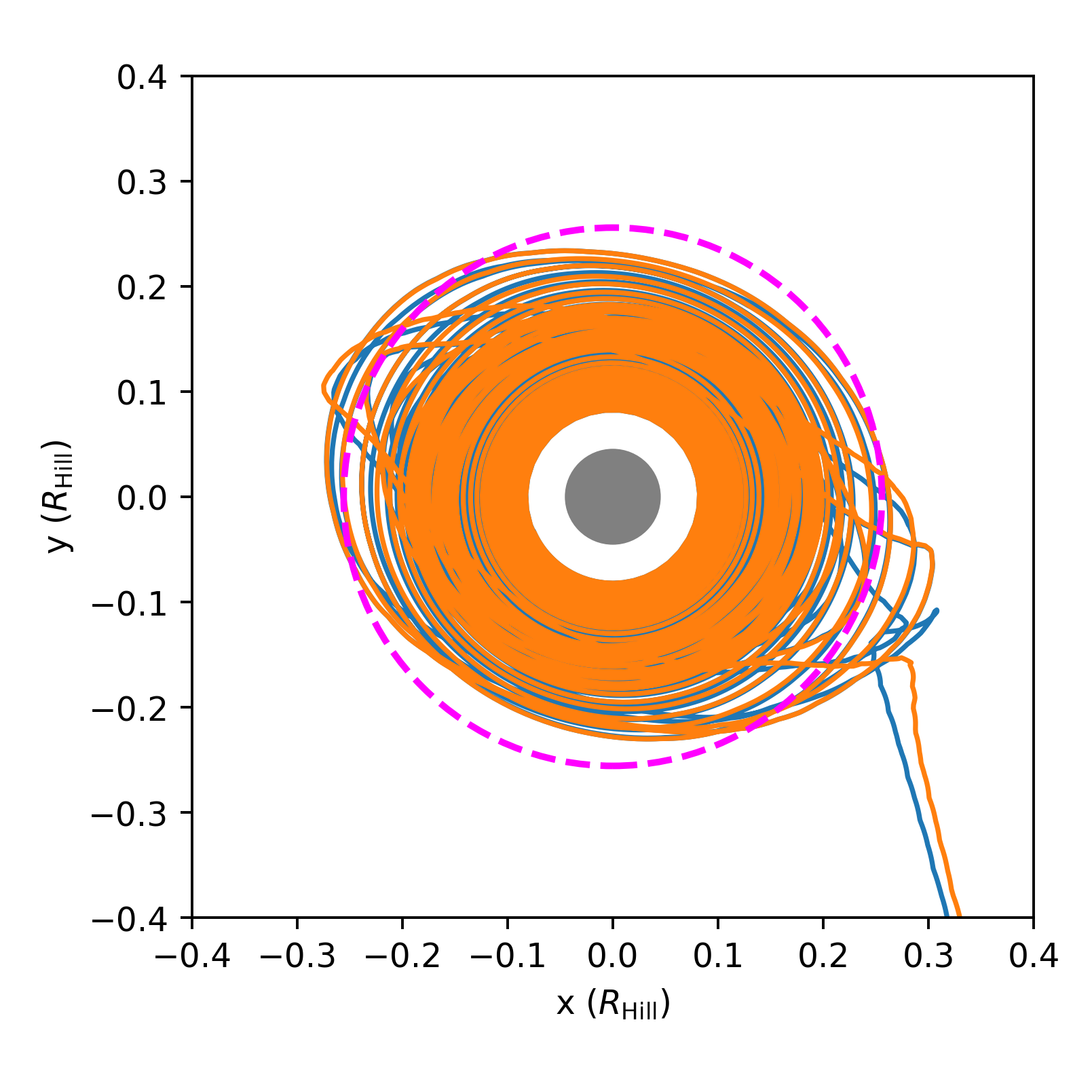}
   \includegraphics[width=0.49\hsize]{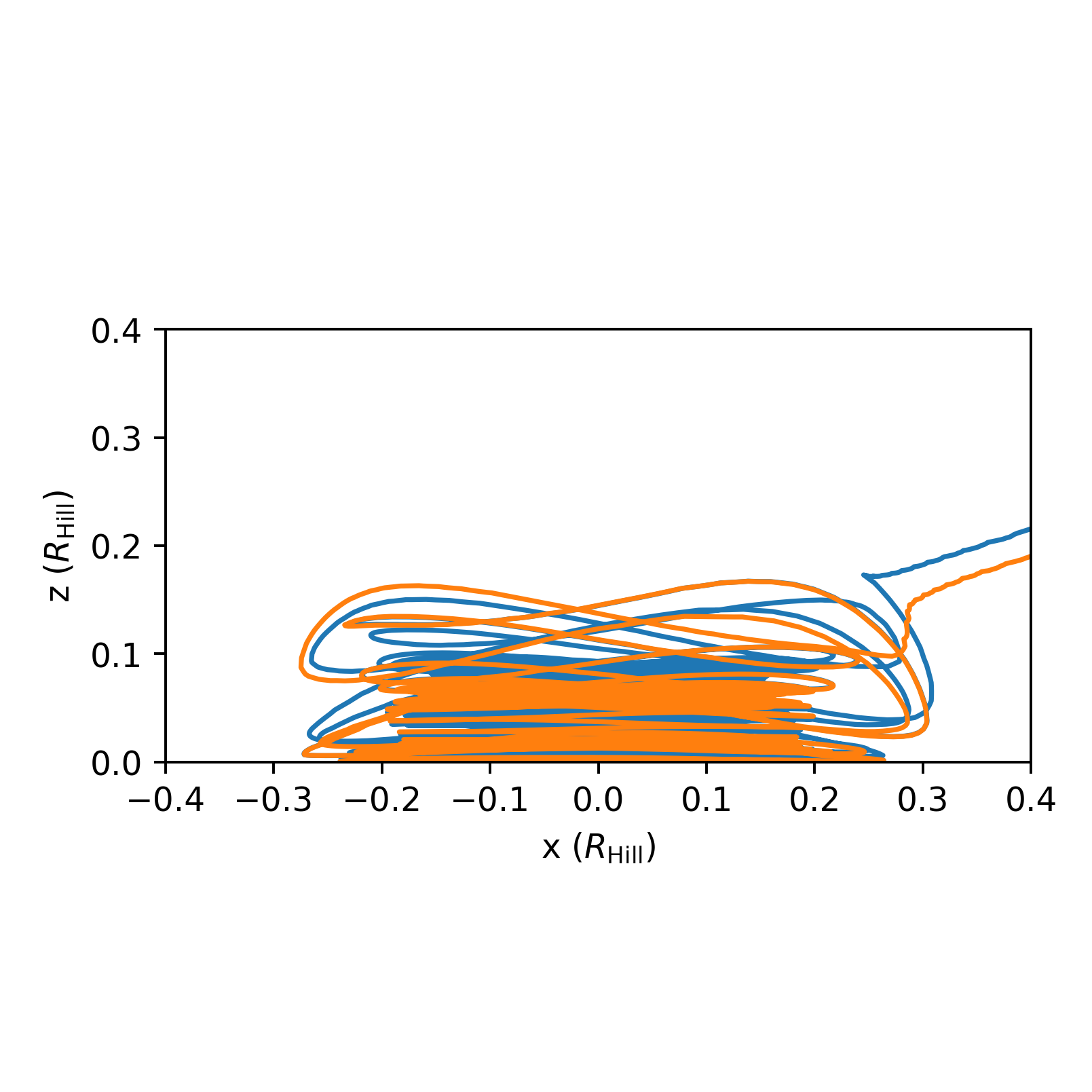}
   \includegraphics[width=0.49\hsize]{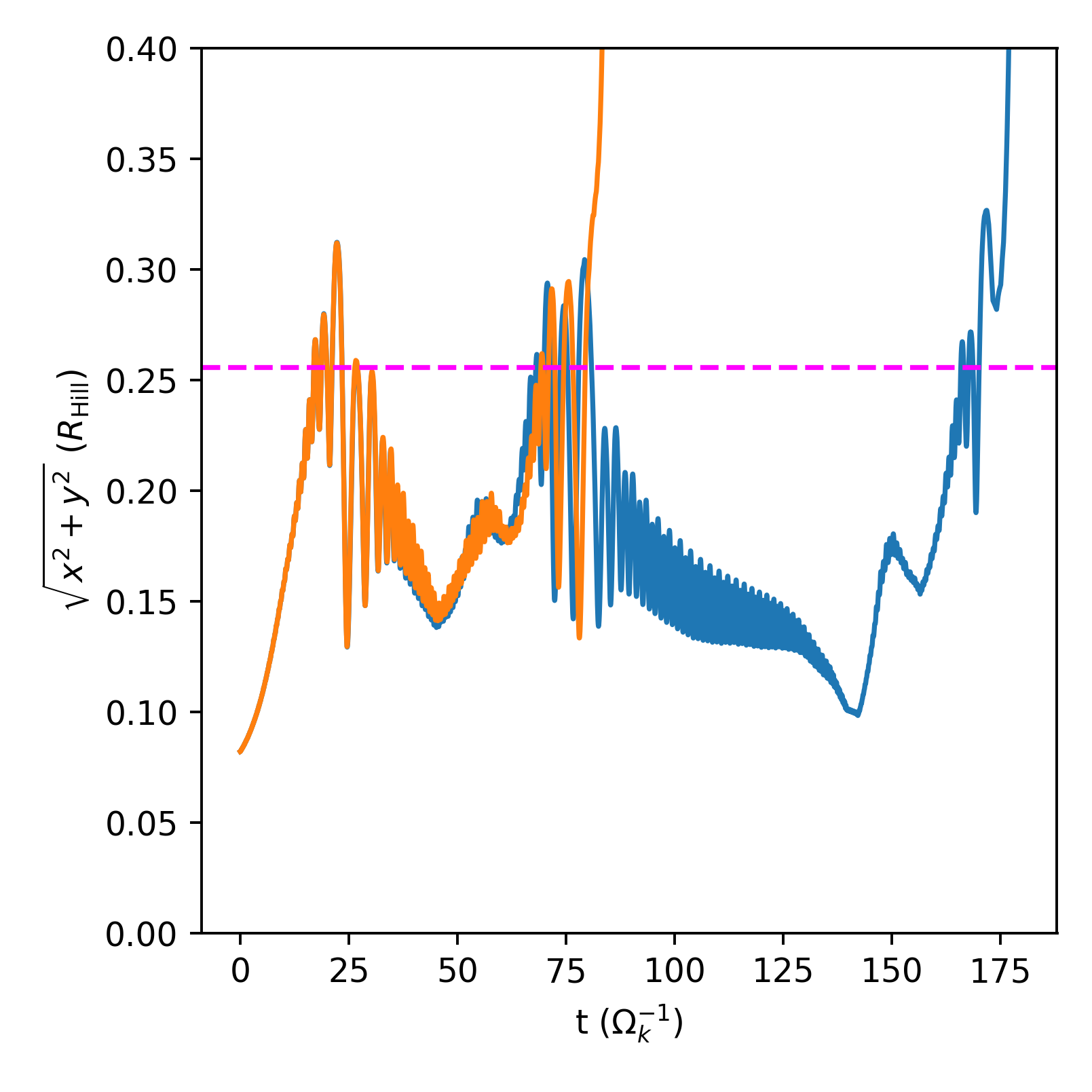}
   \includegraphics[width=0.49\hsize]{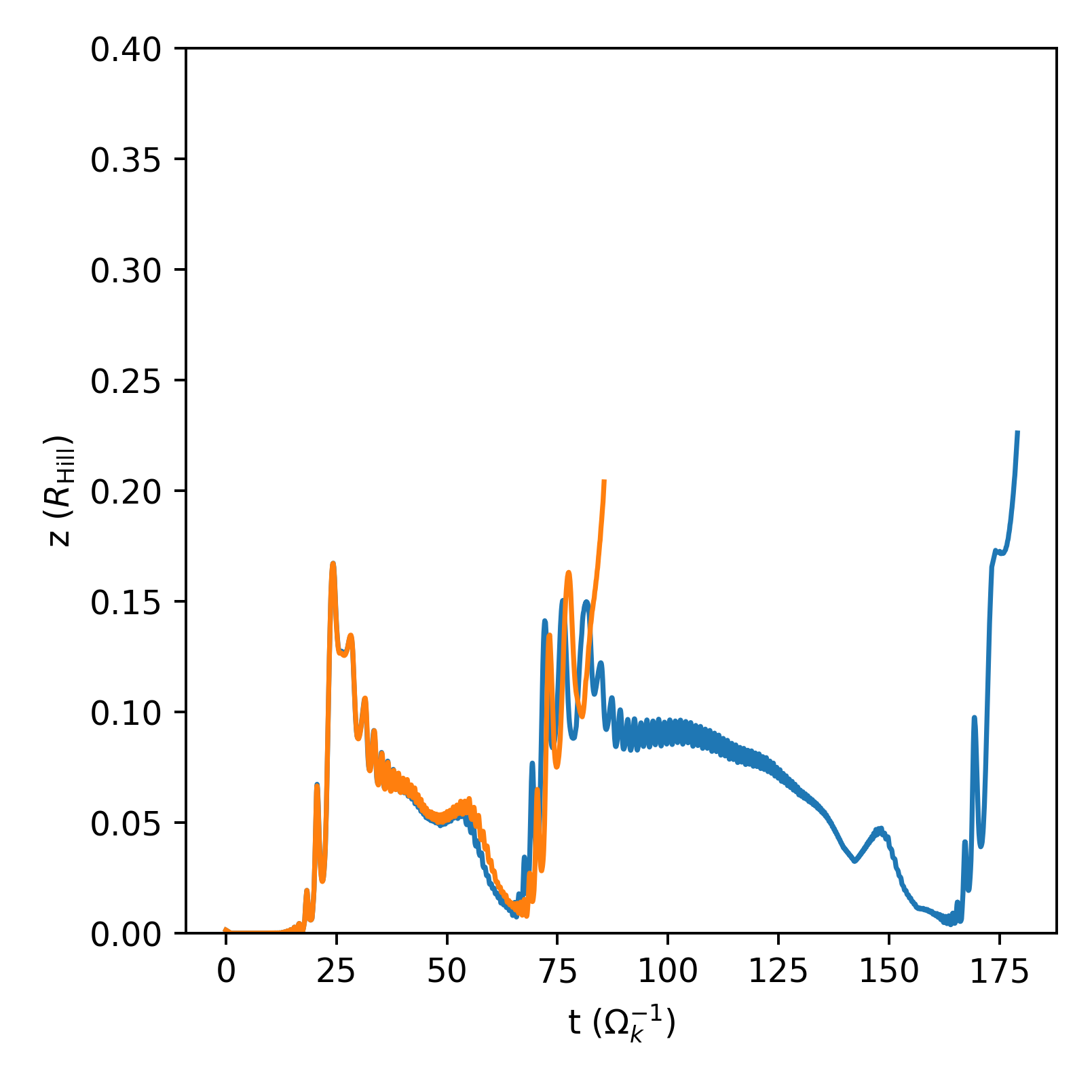}
   \caption{
      Trajectories of two tracer particles integrated forward in time inside the Hill sphere.
      Both particles start in the midplane with the same distance to the protoplanetary core and only slightly different azimuthal angles.
      The starting location was chosen specifically to show the sensitivity of the path to the initial conditions.
      The dashed magenta lines are the atmospheric radius $R_\mathrm{atm}$ as defined in section \ref{sec:length_scales}.
      The planetary core has a radius of $R_\mathrm{core} = 0.044 \, R_\mathrm{Hill}$ and is displayed as a gray circle.
      The white ring in the top left panel is also recycled, but is not part of the trajectory of the two example tracer particles.
      Top left: Midplane view.
      Top right: View of the xz-plane.
      Bottom left: Distance to the z-axis ($\sqrt{x^2+y^2}$).
      Bottom right: Distance to the midplane (z-coordinate).      
      Both tracer particles start to spiral outward in the midplane.
      The tracer particle represented by the orange line escapes the Hill sphere much faster than the particle represented by the blue line, indicating a strong dependence on the initial conditions of the tracer particle.
   }
   \label{fig:trajectory}
\end{figure*}

\begin{figure*}[ht]
   \centering
   \includegraphics[width=0.98\hsize]{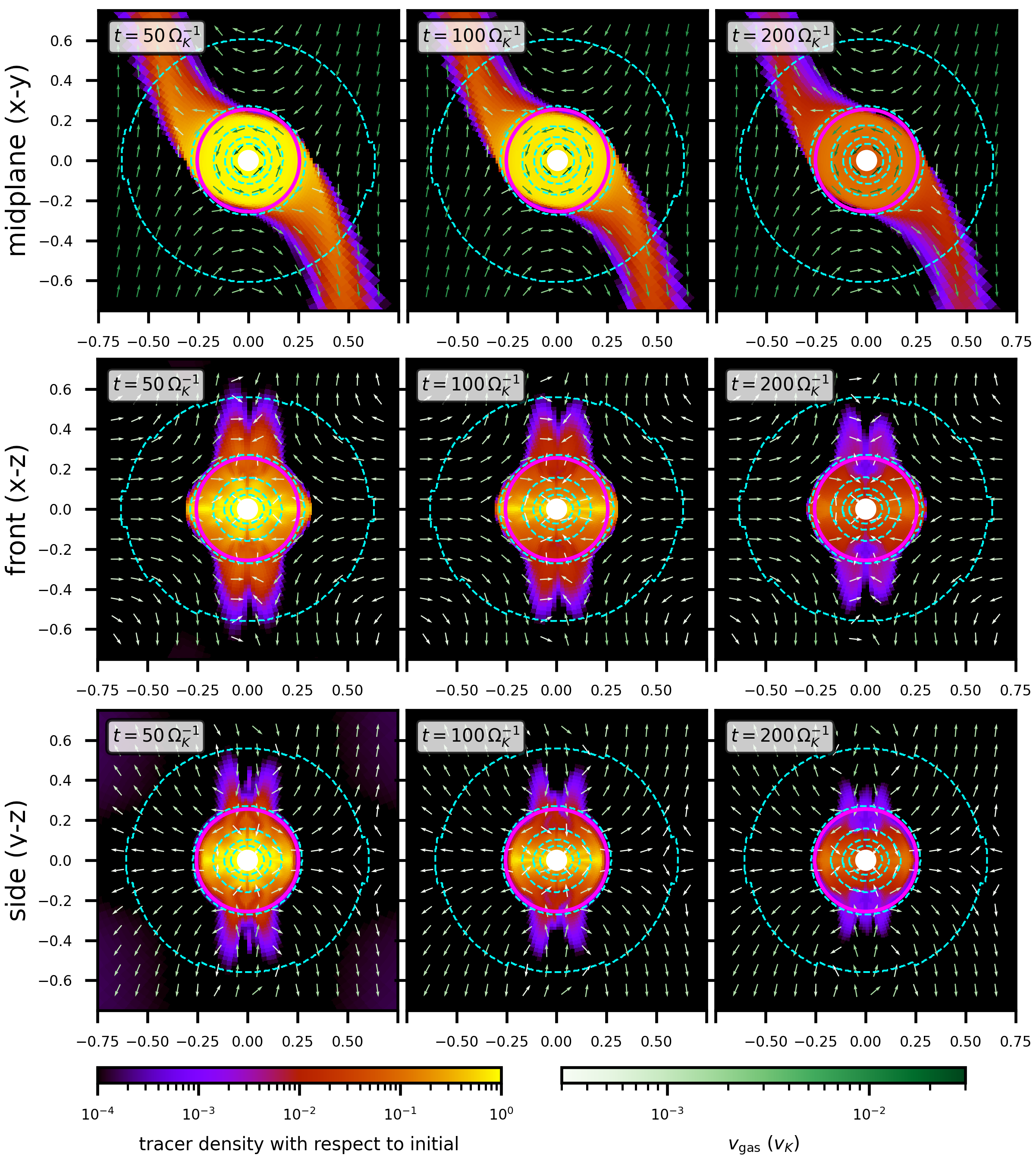}
   \caption{Evolution of the tracer fluid inserted into the Hill sphere of the M1 simulation after thermodynamic equilibrium was reached. At $t=0\,\Omega_K^{-1}$ , the whole Hill sphere is filled with a uniform tracer fluid concentration of unity. The top panel shows the view of the midplane, the middle panel the view of the front, and the bottom panel the side view. The length unit used is $R_\mathrm{Hill}$. The dashed cyan lines are iso-density lines at $[2,4,8,16,32] \, \rho_0$. The magenta circle is the atmospheric radius $R_\mathrm{atm}$, defined such that inside $R_\mathrm{atm}$ , the azimuthal velocity is the largest component everywhere in the midplane, i.e., the gas rotates around the protoplanetary core. The atmosphere is elongated in the x-z view because of the stellar tidal forces.}
   \label{fig:M1_fluid_evolution}%
\end{figure*}

\begin{figure*}[ht]
   \centering
   \includegraphics[width=0.98\hsize]{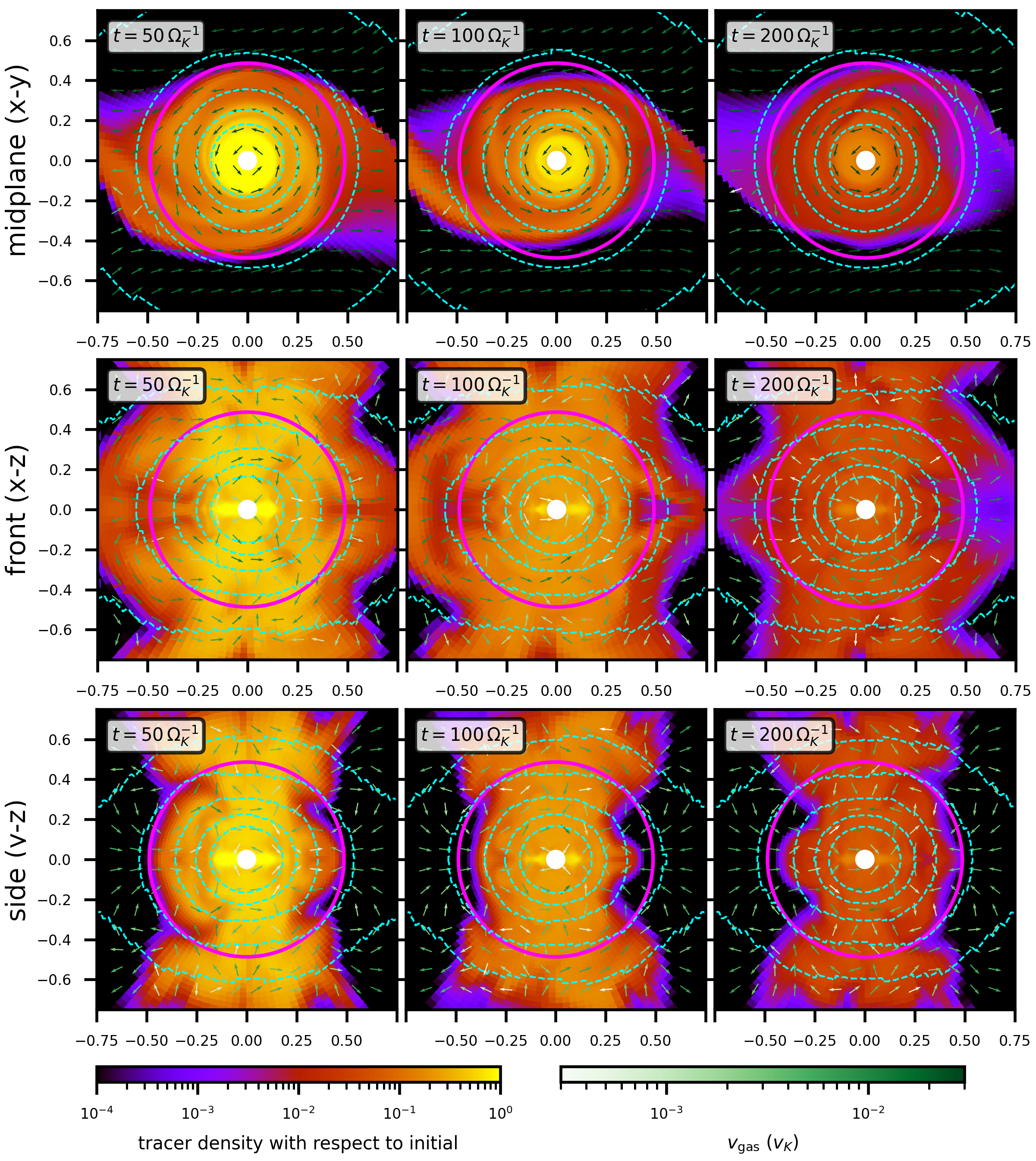}
   \caption{Same as Figure \ref{fig:M1_fluid_evolution}, but for M10. Even though the simulation is in thermodynamic equilibrium, the velocity field changes between snapshots and the atmosphere is asymmetric.}
   \label{fig:M10_fluid_evolution}%
\end{figure*}

\begin{figure*}[ht]
   \centering
   \includegraphics[width=0.98\hsize]{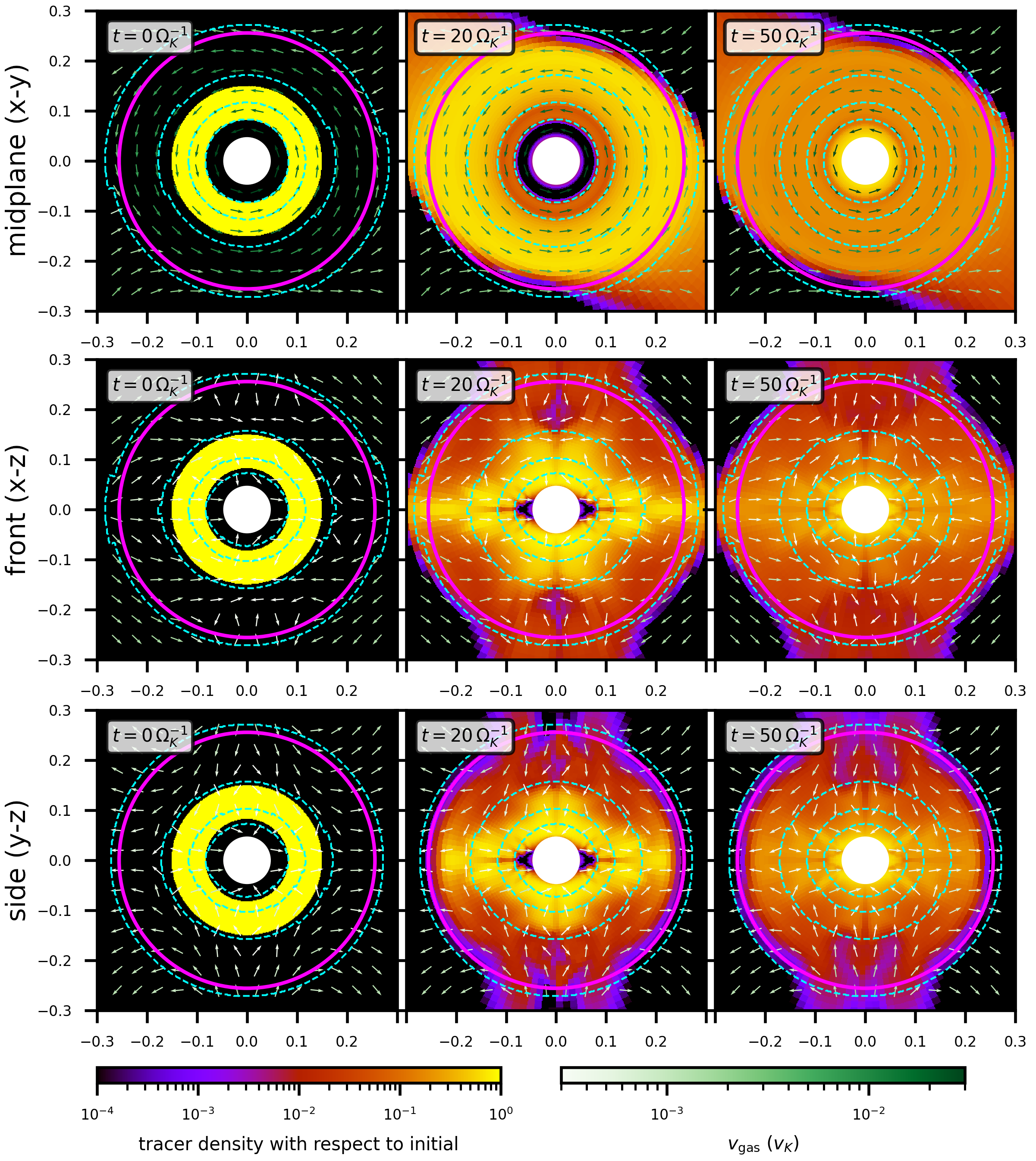}
   \caption{
    Same as Figure \ref{fig:M1_fluid_evolution}, but only a small shell around the core was filled with tracer fluid. The gap between the tracer fluid filled shell and the protoplanetary core highlights how the inner parts of the atmosphere are recycled.
   }
   \label{fig:tracer_M1}%
\end{figure*}

\begin{figure}[ht]
   \centering
   \includegraphics[width=0.98\hsize]{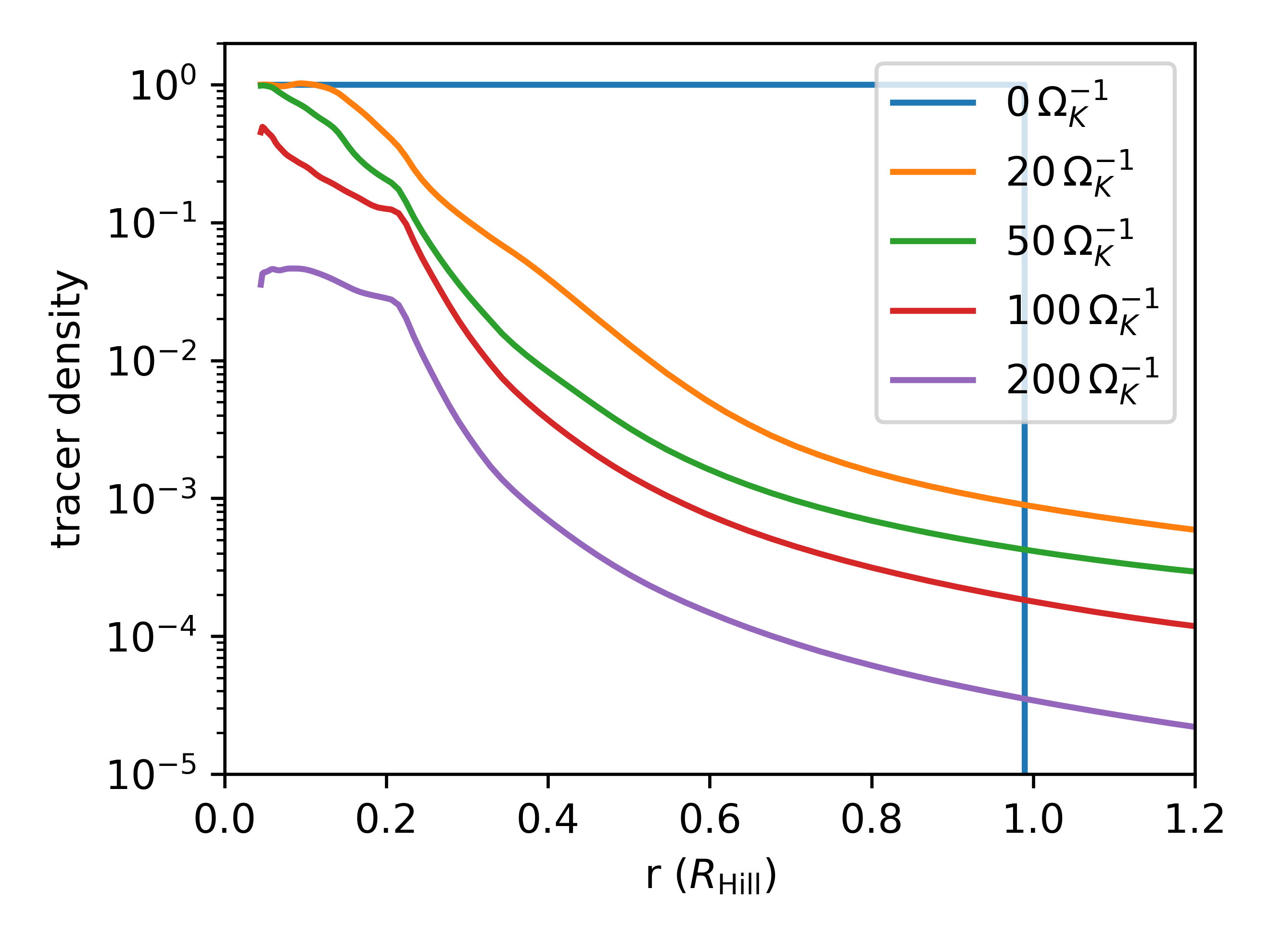}
   \caption{
      Evolution of the 1D-averaged tracer fluid concentration in the Hill sphere of the M1 simulation. The M1 simulation reached thermodynamic equilibrium after $\approx 300 \, \Omega_K^{-1}$ and the tracer fluid was inserted after $2000 \, \Omega_K^{-1}$. After $200 \, \Omega_K^{-1}$ , the tracer fluid concentration decreased to less than $5 \, \%$ even in the inner regions of the protoplanetary atmosphere. This figure shows that the recycling timescale is $\sim$100\,$\Omega_K^{-1}$, explaining that thermodynamic equilibrium is reached after $\sim$100\,$\Omega_K^{-1}$.
   }
   \label{fig:M1_1D_fluid_evolution}%
\end{figure}

\begin{figure*}[ht]
   \centering
   \includegraphics[width=0.98\hsize]{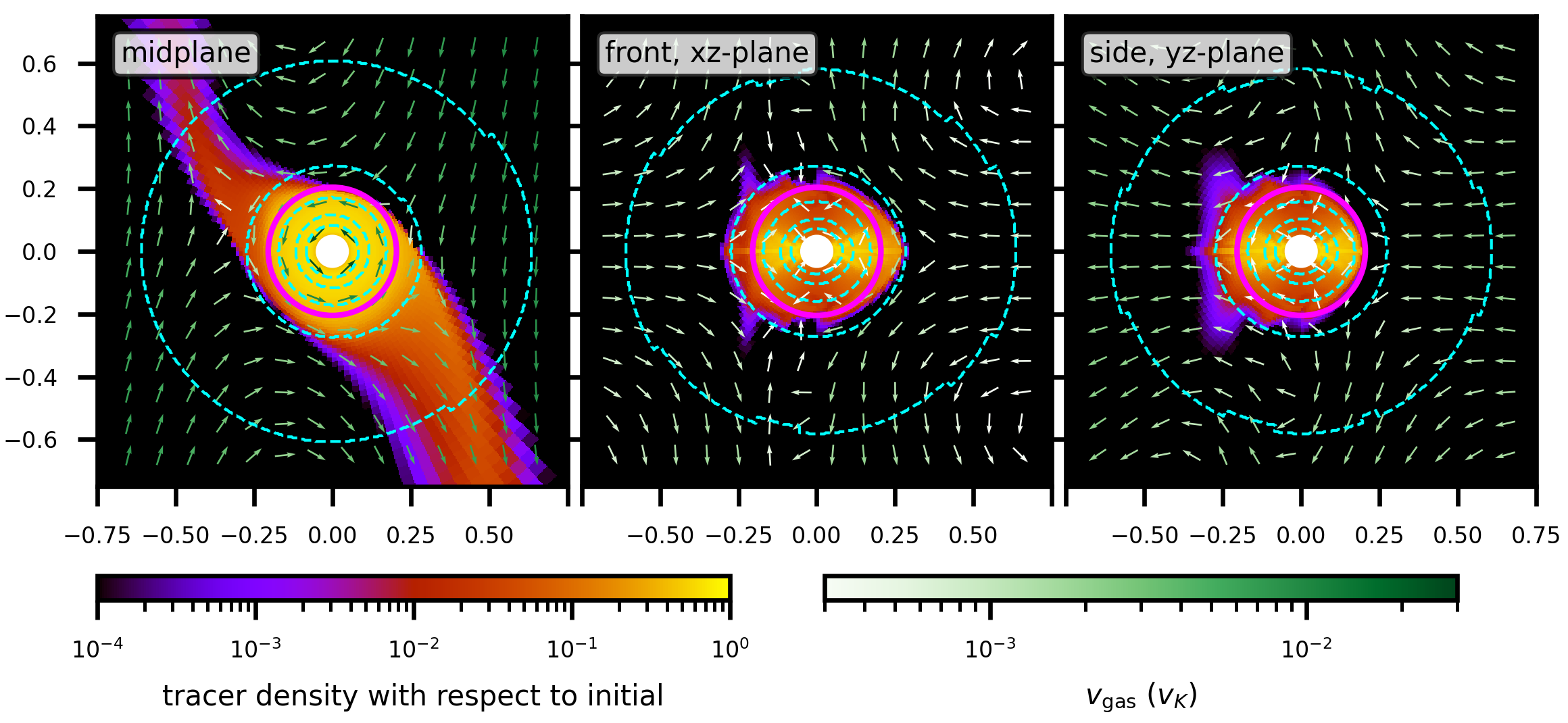}
   \caption{Tracer fluid density at $t= 100 \, \Omega_K^{-1}$ after the tracer fluid was injected into the Hill sphere of the M1-HW simulation after it reached thermodynamic equilibrium. The headwind shifts the horsehoe orbits toward the star, which results in asymmetric recycling streams. Additionally, the headwind recycles the gas directly above and below the protoplanetary atmosphere.}
   \label{fig:tracer_hw}%
\end{figure*}

\begin{figure}[ht]
   \centering
   \includegraphics[width=0.98\hsize]{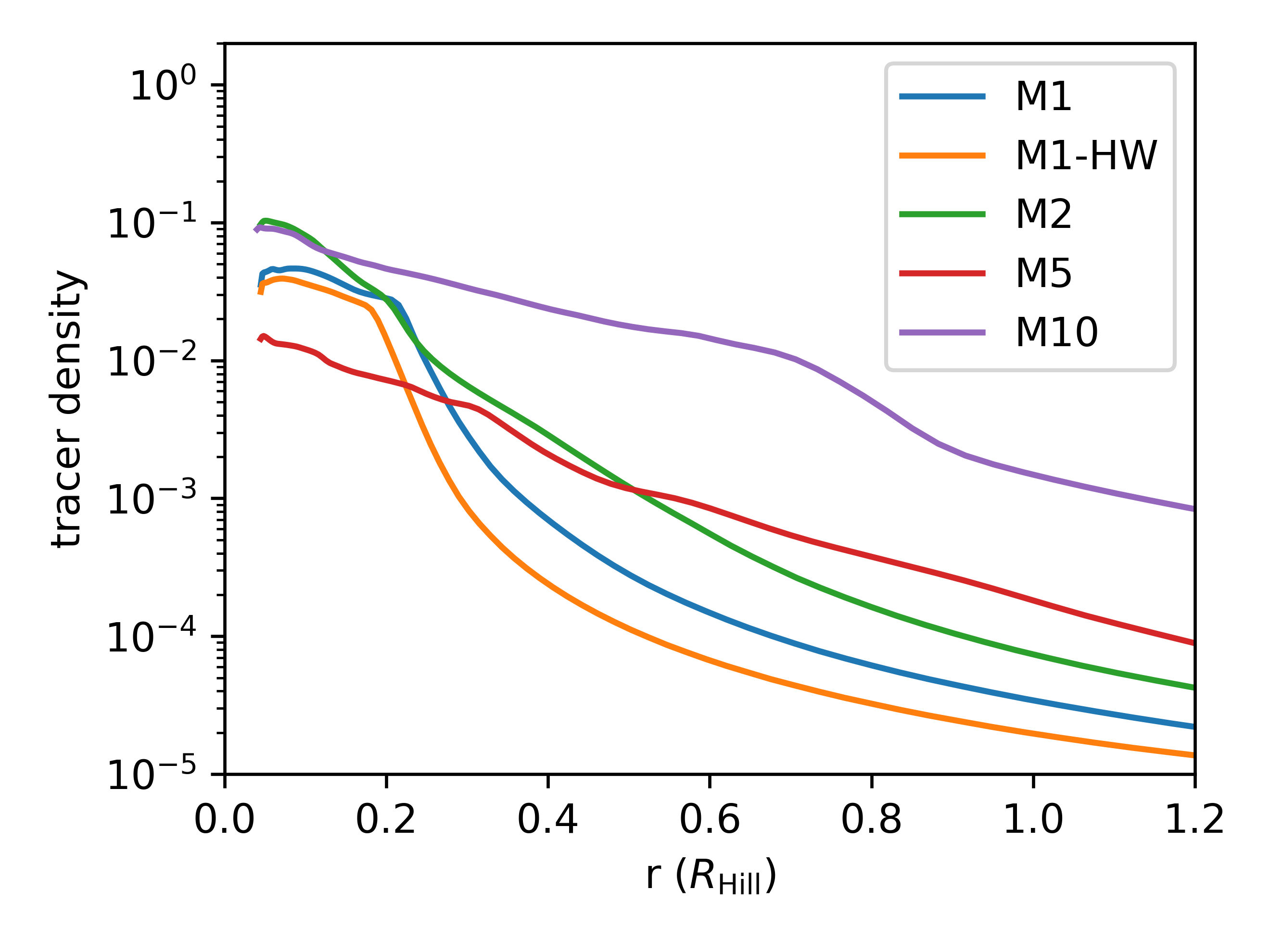}
   \caption{
      1D-averaged tracer fluid concentration for different core masses at $t = 200 \, \Omega_K^{-1}$ after the fluid was introduced into thermodynamic equilibrium.
   }
   \label{fig:1D_fluid_comparison}%
\end{figure}

\begin{figure*}[ht]
   \centering
   \includegraphics[width=0.98\hsize]{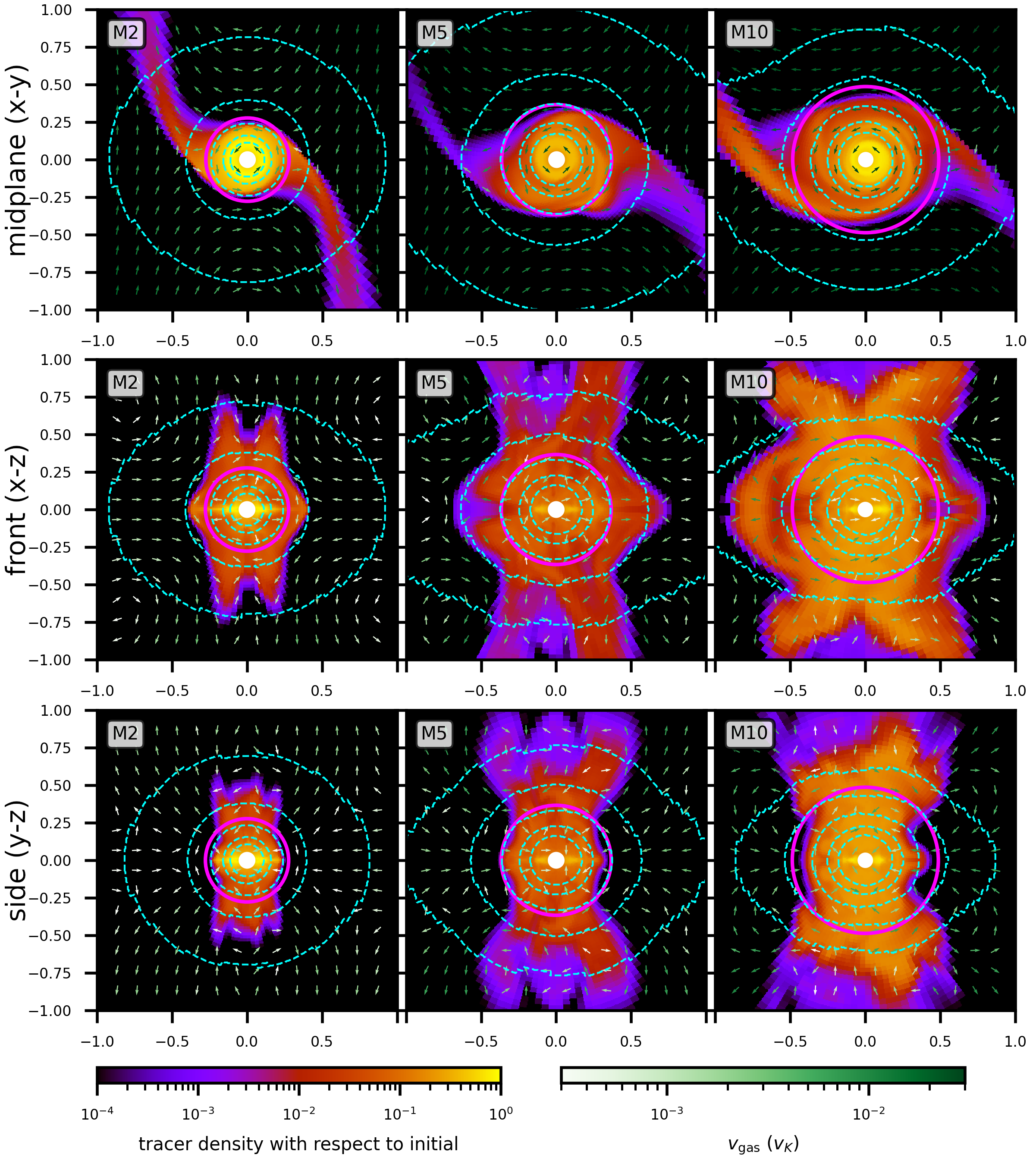}
   \caption{Tracer fluid distribution for simulations with different core masses. After thermodynamic equilibrium is reached, the Hill sphere is initialized with a tracer density of unity and is evolved for an additional $100\, \Omega_K^{-1}$. The length unit for all panels is $R_\mathrm{Hill}$ , which scales with the mass of the core. Because we assumed the same density for all cores, the cores have the same radius in units of $R_\mathrm{Hill}$.
   Top panel: Tracer fluid density in the midplane.
   Bottom panel: Tracer fluid density in the xy-plane seen from the front.}
   \label{fig:tracer_mass}%
\end{figure*}

\begin{figure*}[ht]
   \centering
   \includegraphics[width=0.98\hsize]{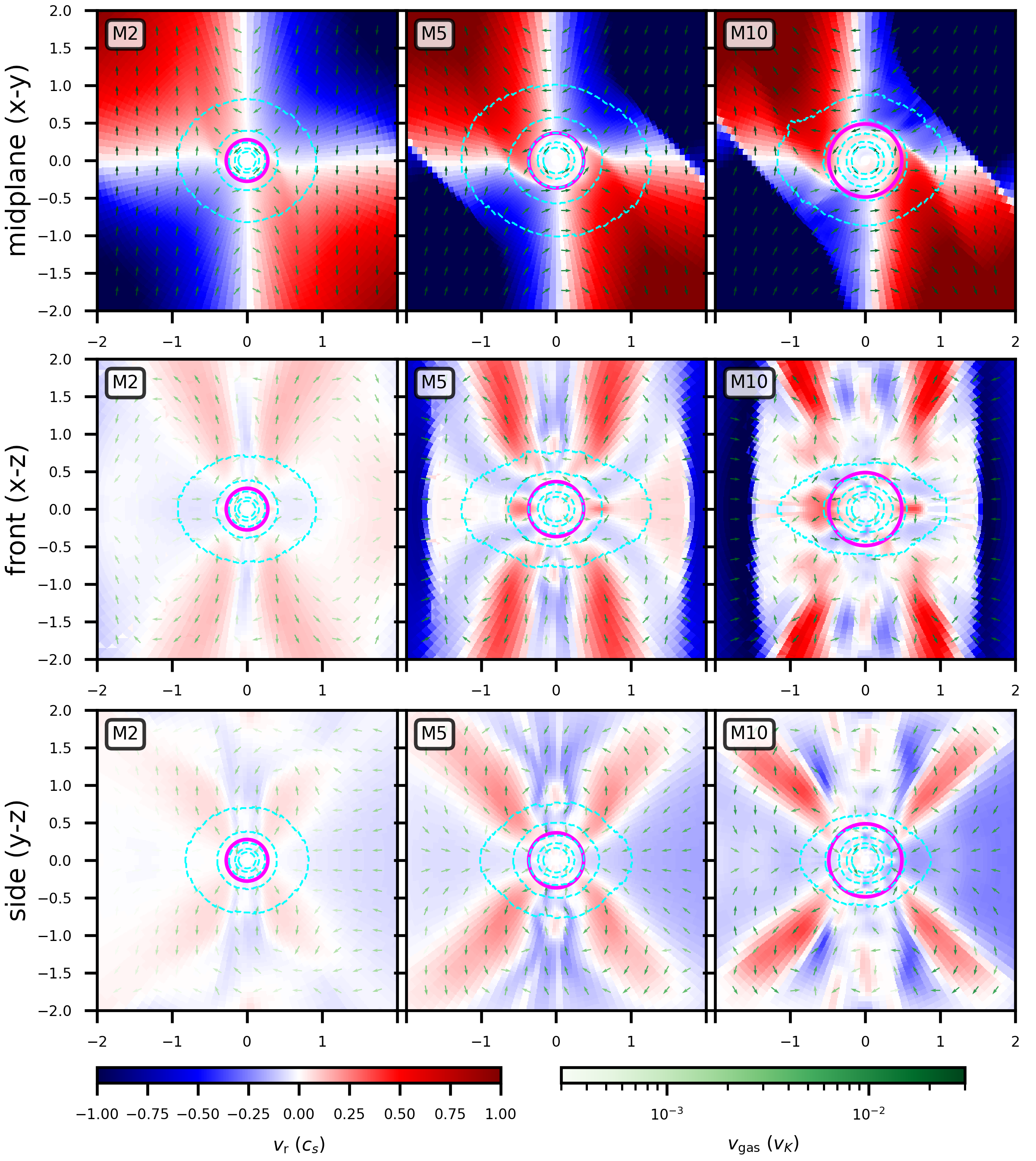}
   \caption{Radial velocity component in units of the local sound speed. The top panel shows the view from the midplane, the middle panel shows the view from the front, and the bottom panel shows the side view. The length unit used is $R_\mathrm{Hill}$. The dashed cyan lines are iso-density lines at $[2,4,8,16,32] \, \rho_0$. The magenta circle is the atmospheric radius $R_\mathrm{atm}$ (see Fig.~\ref{fig:M1_fluid_evolution}).}
   \label{fig:2D_velocity_mass}%
\end{figure*}

\begin{figure*}[ht]
   \centering
   \includegraphics[width=0.98\hsize]{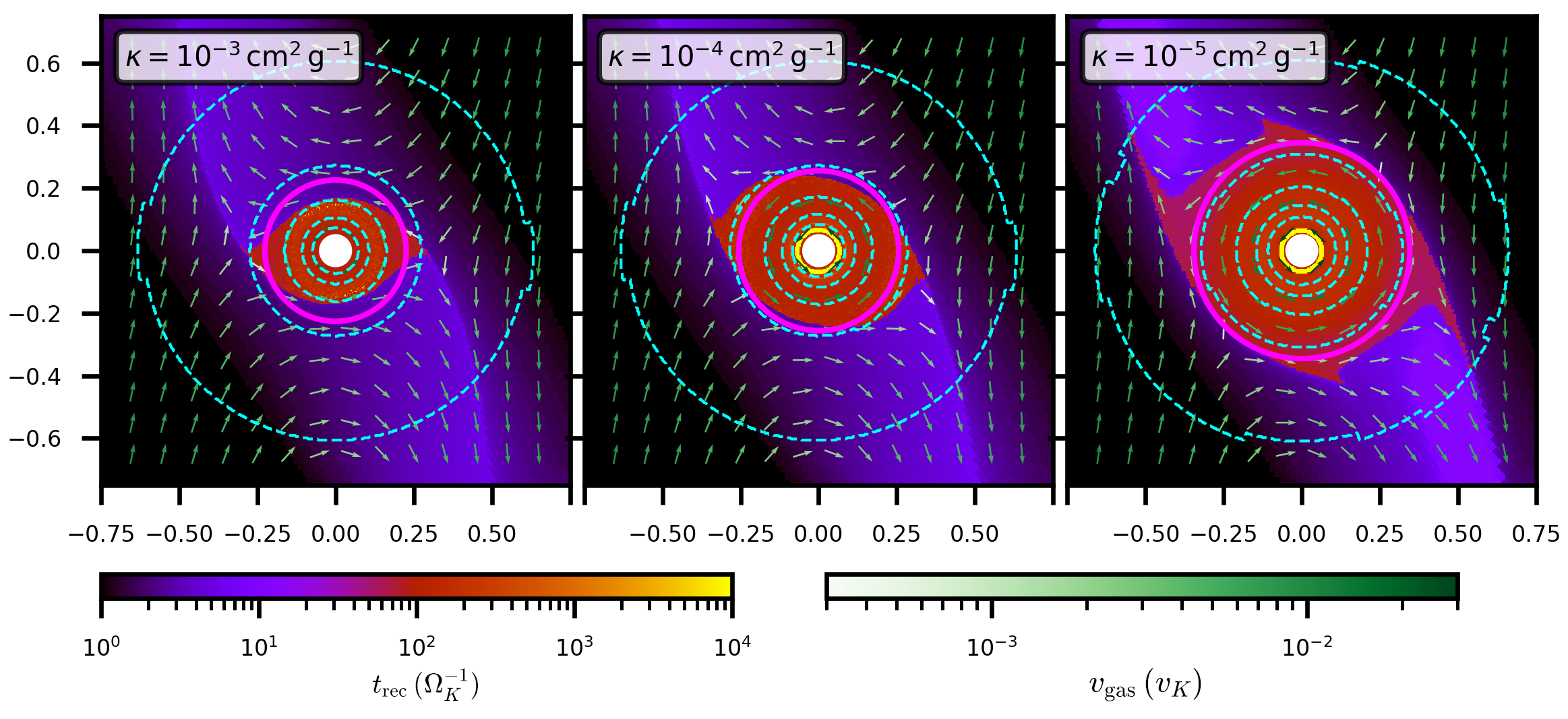}
   \caption{Recycling timescale measured using tracer particles for simulations with different opacities. The center panel is the M1 simulation, which uses our default opacity. After thermodynamic equilibrium is reached, tracer particles are integrated along streamlines backward in time. The measured time until they then escape the Hill sphere is the plotted recycling time. It measures for how long the gas has been inside the Hill sphere.}
   \label{fig:trec_opacity_comparison}%
\end{figure*}

\subsection{Tracer particles}
\label{sec:numerics_tracer_particles}

For our simulations with low planetary mass, $M_c = 1 \, M_\mathrm{Earth}$ , which eventually enter a steady state, we used tracer particles in postprocessing to determine the time the gas has spent inside the Hill sphere to calculate the recycling timescale.
The choice of the Hill sphere to define the recycling timescale is arbitrary.
A change in this radius has little effect on the measured recycling time as long as it exceeds the atmosphere.
The recycling timescale is an additive quantity and orders of magnitude lower outside of the atmosphere than inside.
The Hill sphere was chosen because it is the upper limit at which the planetary core could significantly alter the gas flow.
As tracer particles can easily be integrated along streamlines, they are a computationally inexpensive method for calculating the recycling timescale as a function of space.

After the simulations reached a steady state, we took the velocity field of the simulation averaged over the last 500 outputs at an output frequency of $1 \, \Omega_{K}^{-1}$ and integrated tracer particles along streamlines backward in time.
This removed any residual random oscillations that might disturb the integration of the tracer particles.
We call the time it takes for the tracer particles to escape the Hill sphere the recycling time $t_\mathrm{rec}$.
The recycling time measures the time for which the gas has been in the Hill sphere.
Because of the increased pressure caused by the gravity of the planetary core, atmospheric gas is hotter than circumstellar gas.
This temperature gradient causes the atmosphere to cool through radiative emission.
Therefore, atmospheric gas has a lower entropy than the circumstellar gas.
As atmospheric gas is replenished with fresh circumstellar gas through atmospheric recycling, this entropy loss is compensated for.
Consequently, a longer recycling timescale, that is,~it takes longer for the gas to be replenished, means that gas in this region has had more time to cool radiatively.

We realized that the recycling time strongly depends on the starting position of the tracer particles.
Small spatial variations can cause very different trajectories, such as the number of orbits around the planetary core and thus recycling times.
This is illustrated in more detail in section \ref{sec:result_gas_flow}.
In order to determine the physically meaningful mean value, we arranged the tracer particles on a grid and calculated the median of all tracer particles in a grid cell.
This is justified because the tracer particles trace a gaseous medium that mixes naturally.

For the integration, we used an adaptive fifth-order Runge-Kutta integrator.
The velocity of the tracer particle was calculated from the hydrodynamic velocity field using linear interpolation on a spherical grid.

\subsection{Tracer fluid}
\label{sec:numerics_tracer_fluid}

For our higher core mass simulations, the gas flow structure continued to change even after thermodynamic equilibrium is reached.
Therefore, it is no longer possible to measure the recycling timescale in post processing.
Instead, the recycling timescale has to be measured while the simulation is still running.
While tracer particles could still be used, it is faster and easier to use a tracer fluid instead as the tracer fluid can be advected using the same flux that is calculated for the field of gas mass density.
A tracer fluid is a scalar quantity $\psi$ that is evolved using the advection equation
\begin{align}
    \frac{\partial \psi}{\partial t} + \nabla \cdot (\psi \boldsymbol{v}) &= 0,
\end{align}
where $\boldsymbol{v}$ is the velocity field of the gas.
The density distribution of the tracer fluid $\psi$ is independent of the gas density and can be chosen freely.
However, the tracer fluid density evolves in exactly the same way as the density of the gas, allowing us to trace the motion of the gas without altering it.
For our tracer fluid postprocessing, we evolved the simulation until it reached thermodynamic equilibrium where the temperature of the atmosphere no longer decreases and the density no longer increases.
We then inserted a uniform density tracer fluid in the Hill sphere and continued to evolve the simulation.
The choice of the region that is initially filled with tracer fluid is arbitrary.
We chose the Hill sphere as it serves as an upper estimate of the size of the protoplanetary atmosphere.
The tracer fluid density outside of the Hill sphere, that is, in the circumstellar disk, is zero and no additional tracer fluid enters the simulation domain through the outer boundary.
The tracer fluid may only leave the simulation domain.
After a fixed evolution time, regions that exchange gas with the circumstellar disk on shorter timescales will have a lower concentration of tracer fluid.
The local density of the tracer fluid is therefore a measure for the recycling timescale at that location.

\section{Atmosphere-disk recycling}

First, we revisit the M1 simulation described in \cite{Moldenhauer_2021} and analyze it in even greater detail using additional postprocessing techniques to fully understand the recycling process.
The same postprocessing is used for the other models to perform the parameter study in the next section.
We start by giving an overview of the hydrodynamic quantities in section \ref{sec:result_overview}.
In section \ref{sec:result_equilibrium} we explain the thermodynamic equilibrium that all simulations eventually reach and explain the key differences between low and high core mass cases.
In section \ref{sec:result_gas_flow} we analyze the gas flow in more detail using the trajectories of two sample tracer particles.
In section \ref{sec:result_recycling} we then show the recycling process using a tracer fluid and confirm that the entire atmosphere recycles.

\subsection{Overview}
\label{sec:result_overview}

Figure \ref{fig:sketch} outlines the model and the resulting flow structure in a sketch.
Even though the simulations were performed in spherical coordinates, we used a Cartesian coordinate system to describe the results in the local frame around the planet. 
The $x$- or radial direction points away from the star through the planet, the $y$- or orbital direction points in the orbital direction of the planet, and the $z$- or vertical direction is aligned with the angular momentum vector of the circumstellar disk.
All three coordinates are centered on the planet.
The left panel of figure \ref{fig:M1_energy_density_temperature} displays the 1D-averaged energy budget of the M1 simulation.
The thermal energy of the atmosphere is several orders of magnitude higher than the kinetic energy. The atmosphere is therefore pressure supported.
Inside $R_\mathrm{atm}$ , the kinetic energy density is significantly higher than outside.
This energy is primarily in the azimuthal velocity component, that is, it is from the rotation of the gas around the protoplanetary core.
The density and temperature profiles displayed in the right panel of figure \ref{fig:M1_energy_density_temperature} show a smooth transition from the protoplanetary atmosphere to the circumestellar disk.
Only the velocity profile allows us to distinguish between the atmosphere and the disk.
Hence our definition of $R_\mathrm{atm}$ based on dynamical arguments.

To give an overview of the general flow structure, we first display the midplane density distribution and the velocity field for the M1 simulation in figure \ref{fig:M1_density_evolution}.
After the planetary gravity is switched on, gas accumulates around the planetary core and three distinctive flow regions develop from the initial shearing flow.
Around the planetary core, a protoplanetary atmosphere forms in which the gas flow is dominated by the planetary gravity. In this region, the gas rotates around the planetary core in the azimuthal direction.
Here, up to $0.9 \, R_\mathrm{atm}$, the radial and meridional velocity components do not exceed $5\, \%$ of the total velocity, while at $1.1 \, R_\mathrm{atm}$, for example, the meridional velocity already exceeds $15 \, \%$ of the total velocity.
Along the orbital trajectory of the planet, two horseshoe orbits develop, one in front of the planet, and one behind.
Farther away from the planet in the $x$-direction, the velocity field is dominated by stellar gravity, and the gas in the local frame follows the shearing flow of the disk, that is,~the gas rotates around the central star.
The rotational direction of the protoplanetary atmosphere is against the shearing flow, but is aligned with the horseshoe orbits.
This is due to the conservation of angular momentum, hence the rotational vector of the planetary atmosphere is aligned with the rotational vector of the circumstellar disk.
It becomes clear that the planetary proto-atmosphere and the circumstellar disk are strongly interlinked and have to be studied together.
The protoplanetary atmosphere is not isolated from the circumstellar disk.

\subsection{Thermodynamic equilibrium}
\label{sec:result_equilibrium}

In \cite{Moldenhauer_2021}, we showed that the M1 setup eventually reaches a steady state in which radiative cooling is fully compensated for by atmosphere-disk recycling, that is,~an exchange of entropy between the atmosphere and the circumstellar disk by gas advection.
Radiative cooling transports energy outward and recycling of low-entropy atmospheric gas with high-entropy circumstellar gas transports energy inward, while the net mass flux across shells effectively vanishes.
The cooling timescale itself is not measured, but the temperature and density are observed to eventually become constant in time, indicating that radiative cooling is fully compensated for by atmospheric recycling.
This shows that the cooling timescale is longer than the recycling timescale despite the very low opacity, highlighting the efficiency of the recycling process.
The simulations in this paper explore a wide range of core masses and a scenario with and without headwind.
All our simulations eventually reached thermodynamic equilibrium after $\approx 300 \, \Omega_K^{-1}$ , which is on the same order of magnitude as their respective recycling timescales.
In thermodynamic equilibrium, the atmosphere does not continue to cool, despite a continuous loss of radiation energy.
Only smaller random oscillations remain, which tend to be stronger for higher core masses.
For all simulations, the spherically averaged temperature oscillates by less than $1\,\%$ and the spherically averaged density by less than $2\,\%$.
This thermodynamic equilibrium does not necessarily mean that the flow structure needs to be constant in time.
For higher core masses, we observe that the atmosphere starts to become turbulent.
Here, the velocity field changes in time, but the simulations stay in thermodynamic equilibrium.

Additionally, we were able to remove the oscillations caused by the initial collapse of the atmosphere that were reflected by the radial boundaries.
To remove the oscillations, we ran the simulations until they reached equilibrium and then averaged over several hundred output files at an output frequency of $1\,\Omega_K^{-1}$.
We then continued the simulation from this averaged state.
The turbulent flow observed for higher core masses is independent of these oscillations.
Removal of these oscillations did not change the equilibrium temperature or the recycling time of the atmosphere.
However, for maximum accuracy, we performed all postprocessing steps from this averaged state. 

\subsection{Characteristics of the gas flow}
\label{sec:result_gas_flow}

In order for atmospheric gas to recycle, there have to be regions in which the atmosphere accretes material and regions in which material leaves the atmosphere, that is,~a continuous exchange of gas between the atmosphere and the circumstellar disk.
Figure \ref{fig:trajectory} displays the trajectories of two tracer particles that start in the midplane close to the planetary core at almost the exact same position.
Both tracer particles start with the same distance to the core, they only differ slightly in their initial azimuthal coordinate.
At first, both particles spiral outward while being stuck in the midplane, that is,~$z \approx 0$.
After they reach a distance to the center of the core $\sqrt{r^2-z^2} \approx R_\mathrm{atm}$, they reach the zone where the gas from the protoplanetary atmosphere starts to mix with the circumstellar gas from the horseshoe orbits.
Here, they drift away from the midplane and rapidly gain altitude before they slowly move back toward the midplane.
During this vertical mixing process, the particles oscillate around a distance of $\sqrt{r^2-z^2} \approx R_\mathrm{atm}$ from the center of the vertical axis.
The process then repeats and the particles again move away from the midplane.
The particle represented by the orange trajectory is then picked up by one of the horseshoe orbits and is removed from the Hill sphere, while the particle represented by the blue trajectory repeats the drift toward the midplane for a second time before it is finally ejected.
This illustrates how sensitive the trajectories of the tracer particles are to the initial conditions.
The boundary between the protoplanetary atmosphere and the horseshoe orbits features an irregular flow pattern causing continuous mixing between the atmospheric and circumstellar gas.
For previous 2D simulations \citep{Ormel_2015}, a well-defined separatrix existed that separated the atmosphere from the circumstellar flow.
In 3D the topology is different and this clear distinction is no longer possible.

\subsection{Recycling of the entire atmosphere and turbulence}
\label{sec:result_recycling}

Figure \ref{fig:M1_fluid_evolution} shows the tracer fluid concentration at different points in time for the M1 simulation.
The tracer fluid was introduced with a concentration of unity inside the Hill sphere after the simulation reached thermodynamic equilibrium.
The protoplanetary atmosphere is slightly elongated in the x-direction, as can be seen when the front view is compared to the side view.
This is because of the stellar tidal forces that stretch the atmosphere along the radial direction.
Except for directly above and below the planetary core, the region outside of $R_\mathrm{atm}$ is cleared from the tracer fluid more than 2 orders of magnitude faster than inside, indicating that this region is not part of the atmosphere.
After only $50 \, \Omega_K^{-1}$ , the tracer fluid concentration in this region is lower than $10^{-4}$, except for the horseshoe region, where the tracer fluid from the atmosphere is transported away.
This observation agrees with the structure of the velocity field because only for $r<R_\mathrm{atm}$ does the gas rotate around the protoplanetary core.
This motivates our definition of $R_\mathrm{atm}$ from section \ref{sec:length_scales} as it marks the transition from short recycling to significantly longer recycling.

The high-mass case is depicted in figure \ref{fig:M10_fluid_evolution}, which shows the time evolution of the tracer fluid for the M10 simulation.
Here, the protoplanetary atmosphere is asymmetric compared to the low-mass case.
The asymmetries arise from turbulence, which can be observed by the changing velocity field especially in the front and side view.
However, as the turbulence occurs on a subsonic timescale, it is only visible in the velocity field.
Even the turbulent simulations eventually enter thermodynamic equilibrium where density and temperature stay approximately constant in time.
When we compared the displayed tracer fluid densities to results where we artificially kept the velocity field constant after the tracer fluid was injected, we observed that for higher core masses, the resulting tracer fluid density was significantly different.
This is because the random changes in the velocity field for the high-mass case have a significant impact on the recycling process.
Turbulence affects the recycling process in two ways: At the boundary layer between the atmospheric flow around the planetary core and the horseshoe orbits, it increases mixing.
Additionally, turbulence increases the mixing inside the atmosphere, that is, it transports low-entropy gas toward the boundary layer and recycles high-entropy gas from the boundary layer to the inner parts of the atmosphere.
The addition of viscosity to our inviscid simulations might soften or remove the observed turbulent flow and provide additional heating.

In Figure \ref{fig:tracer_M1} we highlight how the gas is distributed inside the inner parts of the M1 simulation.
First, we waited for the simulation to enter thermodynamic equilibrium, then we filled a small spherical shell around the planetary core with the tracer fluid.
We intentionally left a small gap between the spherical shell filled with tracer fluid and the planetary core.
This allowed us to show that even the innermost regions of the atmosphere are recycled efficiently.
After the introduction of the tracer fluid, we let the simulation continue to evolve in its thermodynamic equilibrium.
The tracer fluid was quickly distributed throughout the entire protoplanetary atmosphere on a timescale of $t_\mathrm{mix} \sim 10 \, \Omega_K^{-1}$, including the region between the initial tracer fluid location and the core.
The inner region that was initially left without tracer fluid was eventually filled with tracer fluid primarily through the vertical direction.
The midplane region of the inner atmospheric parts took longest to be filled with tracer fluid as this is the region in which gas is being transported outwards.
At the boundary between the horseshoe orbits and the atmosphere, some of the tracer fluid mixed with the gas of the horseshoe orbit and was transported away from the planet and back into the circumstellar disk.
This process slowly decreased the tracer fluid concentration in the atmosphere, showing how the gas is being recycled on a timescale of $t_\mathrm{rec} \sim 100 \, \Omega_K^{-1}$.
In addition to the gas exchange through the horseshoe orbits, gas was accreted along the vertical axis, indicated by the lower tracer fluid concentration along the vertical axis compared to the midplane.
In agreement with the tracer particle trajectories, the tracer fluid shows that the gas inside the protoplanetary atmosphere is constantly mixed and gas is exchanged with the circumstellar disk through the horseshoe orbits and through vertical infall.

In Figure \ref{fig:M1_1D_fluid_evolution} we show the evolution of the 1D-averaged tracer fluid concentration inside the Hill sphere for the M1 simulation.
Initially, all cells that are fully within the Hill sphere were filled with a tracer fluid.
When we then continued to evolve the simulation, the tracer fluid concentration started to decrease.
While the concentration in the inner regions, $r<R_\mathrm{atm}$, decreased more slowly, all regions experience a depletion of the tracer fluid on a recycling timescale $t_\mathrm{rec} \sim 100 \, \Omega_K^{-1}$.
After $200 \, \Omega_K^{-1}$ , the tracer fluid concentration decreased to lower than $5 \, \%$ everywhere within in the protoplanetary atmosphere.
This shows that even the atmospheric region in which the gas rotates around the protoplanetary core is recycled relatively quickly.
Additionally, this explains why it takes $\sim 100 \, \Omega_K^{-1}$ for the simulation to reach thermodynamic equilibrium. 

\section{Parameter study}
\label{sec:parameter_study}

To understand the parameters that determine the efficacy of atmospheric recycling, we conducted a parameter study.
First, we analyze the effect of the headwind that the protoplanet experiences in section \ref{sec:result_hw}.
In section \ref{sec:result_mass}, we then analyze the influence of the planetary core mass on the structure of the atmosphere and on the resulting recycling timescale.
Because the boundaries have improved and new postprocessing techniques are available, we revisit the claim of \cite{Cimerman_2017} that opacity has little effect and determine its influence on the recycling timescale.
Finally, section \ref{sec:atmospheric_mass} investigates the influence of all studied parameters on the final atmospheric mass in thermodynamic equilibrium.

\subsection{Recycling with headwind}
\label{sec:result_hw}

Including the headwind introduces an asymmetry: The horseshoe orbits are no longer centered on the protoplanet, but are shifted toward the star.
The gas that rotates around the star with the same frequency as the planet has a smaller separation from the star.
This asymmetry allows the horseshoe orbits to penetrate deeper into the protoplanetary atmosphere, potentially increasing the recycling rate.
Figure \ref{fig:tracer_hw} shows the evolution of the tracer fluid for the M1-HW simulation.
As expected, the recycling streams are no longer symmetric.
Compared to the simulation without headwind, more tracer fluid leaves the protoplanetary atmosphere behind the planet, but in turn, less tracer fluid leaves the atmosphere in front of the planet.
This is a result of the shift of the horseshoe orbits toward the star.
Additionally, the headwind causes the regions of slow recycling along the vertical axis to disappear.
Gas that was previously able to accumulate along the vertical axis is now swept away by the headwind.
Instead, this region also moved toward the star, as is visible from the side view of the protoplanetary atmosphere.
The density structure is not affected as it is dominated by the core gravity instead of the flow structure.
These effects result in a generally slightly smaller protoplanetary atmosphere.
However, even though we overestimated the headwind magnitude by roughly a factor of 5, the effect on the size of the atmosphere is relatively small.
For higher core masses and therefore higher thermal masses, the effects of the headwind become even weaker.

\subsection{Core mass comparison}
\label{sec:result_mass}

In section \ref{sec:result_recycling} we have explained one key difference between lower core masses and higher core masses: for higher core masses, the velocity field changes in time even after thermodynamic equilibrium was reached.
In this section, we explore the effects of the core mass on the recycling of the protoplanetary atmosphere.
We waited until our simulations with core masses $M_c \in [2,5,10]$ reached thermodynamic equilibrium, where the density and temperature distribution of the atmosphere become constant in time.
While thermodynamic equilibrium was reached after $\approx 300 \, \Omega_K^{-1}$, we simulated for $1000 \, \Omega_K^{-1}$ before injecting the tracer fluid in the Hill sphere.

Figure \ref{fig:tracer_mass} displays the tracer fluid concentration at $t=100 \, \Omega_K^{-1}$ after it was injected for our simulations with core masses $M_c \in [2,5,10] \, M_\mathrm{Earth}$.
The length scale used is $R_\mathrm{Hill}$ , which increases with the mass of the core.
Because we assumed the same planetary core density for all core masses, the size of the core is identical and measured in $R_\mathrm{Hill}$ across all simulations: $R_\mathrm{core} = 0.044 \, R_\mathrm{Hill}$.
However, even with this variable scaling, the displayed atmosphere increases with core mass.
This means that the size of the atmosphere increases faster than the radius of the core and the Hill sphere when the core mass is increased.
The midplane view of figure \ref{fig:tracer_mass} shows that as the core mass increases, more structure develops.
The front and side views also show that the protoplanetary atmosphere becomes increasingly asymmetric for higher core masses.
These asymmetries are not constant in time but change continuously as the simulation progresses and are caused by the onset of turbulence in the high-mass case.
However, because the velocity field changes occur on a subsonic timescale, the temperature and density profile stay approximately constant in time.
Although the atmosphere for higher core masses is turbulent, the protoplanetary atmosphere remains in thermodynamic equilibrium.
Additionally, the larger atmosphere causes the tidal effects to be more pronounced, for higher core masses the atmosphere is significantly more stretched in the x-direction compared lower core masses.
In the high-mass case, the density distribution appears less spherical and is squished in the vertical direction.
However, this is mainly due to the larger size of the atmosphere, in which the vertical stratification of the circumstellar density profile has a stronger effect.
The protoplanetary atmosphere itself still remains more spherical.

Figure \ref{fig:2D_velocity_mass} shows the radial velocity component for the different core masses as a 2D field in units of the local sound speed.
At higher core masses, the radial in- and outflow becomes stronger because the gravitational acceleration by the planetary core is stronger.
For higher core masses, $M_c > 2 \, M_\mathrm{Earth}$ accretion shocks develop outside of the Hill sphere.
However, as the atmosphere is significantly smaller than the Hill sphere, the whole recycling process occurs at subsonic velocities.
Additionally, for higher core masses, the radial velocity also becomes increasingly asymmetric.
This is because of the turbulent flow in and around the protoplanetary atmosphere at higher core masses.

For all core masses we analyzed, the azimuthal velocity component is sub-Keplerian at less than $50\,\%$ of the Keplerian velocity with respect to the planetary gravity.
This shows that the atmosphere is mainly pressure supported, regardless of the core mass.
Compared to the low-mass case, the high-mass case results in a flatter rotational velocity profile.
This is caused by the more turbulent atmosphere at higher core masses.
Previous studies also found a mainly pressure-supported atmosphere \citep{Moldenhauer_2021, Cimerman_2017}.

\subsection{Opacity comparison}
\label{sec:result_opacity}

For our simulations we deliberately chose a very low opacity, $\kappa = 10^{-4} \, \mathrm{cm}^2~\mathrm{g}^{-1}$, in order to keep the cooling timescale short and therefore save on computation time.
This raises the question whether this decreased opacity affects the recycling timescale.
Additionally, we did not include self-gravity and the opacity was constant, therefore changing the opacity has the same effect on the flow structure as changing the ambient density of the disk.
Here, we explore simulations with $\kappa = 10^{-3} \, \mathrm{cm}^2~\mathrm{g}^{-1}$ and $\kappa = 10^{-5} \, \mathrm{cm}^2~\mathrm{g}^{-1}$.
While these opacities are still considerably lower than what would be realistically expected, they serve as a good comparison to verify the effect that opacity has on the recycling timescale.
All our simulations start optically thick and only slightly increase their optical depth when thermodynamic equilibrium is reached.
In Figure \ref{fig:trec_opacity_comparison} we display the recycling time measured using tracer particles for the different opacities.
Other parameters such as the boundary temperature or the disk velocity profile remain unchanged by this change in opacity.
The size of the atmosphere increases when the opacity is decreased, which at first glance seems counterintuitive.
In the regime of optically thick atmospheres, a lower opacity results in a higher radiative flux and in an increase in cooling efficiency; the atmosphere therefore contracts more.
This results in a higher pressure gradient, which in turn allows for a larger pressure-supported atmosphere.
However, as Figure \ref{fig:trec_opacity_comparison} shows, the recycling timescale remains unchanged.
The opacity only affects the local cooling timescale, while the recycling timescale is affected by the core mass and probably by the global dynamical timescale at the location of planet.
The cooling timescale and the recycling timescale are therefore independent of each other.
Because of our artificially low opacity and the result that the recycling timescale is independent of the cooling timescale, we can conclude that at realistic opacities recycling will also halt the contraction of the atmosphere at the explored parameter space.

The rotational velocity of the atmosphere significantly increases when the opacity is decreased.
This is because of the contraction of the lower entropy atmosphere and the conservation of angular momentum.
However, outside of the protoplanetary atmosphere toward the circumstellar shearing flow, the opacity has virtually no effect on the gas velocity.
As expected, the opacity only changes the local behavior of the atmosphere.

\begin{figure}[ht]
   \centering
   \includegraphics[width=0.98\hsize]{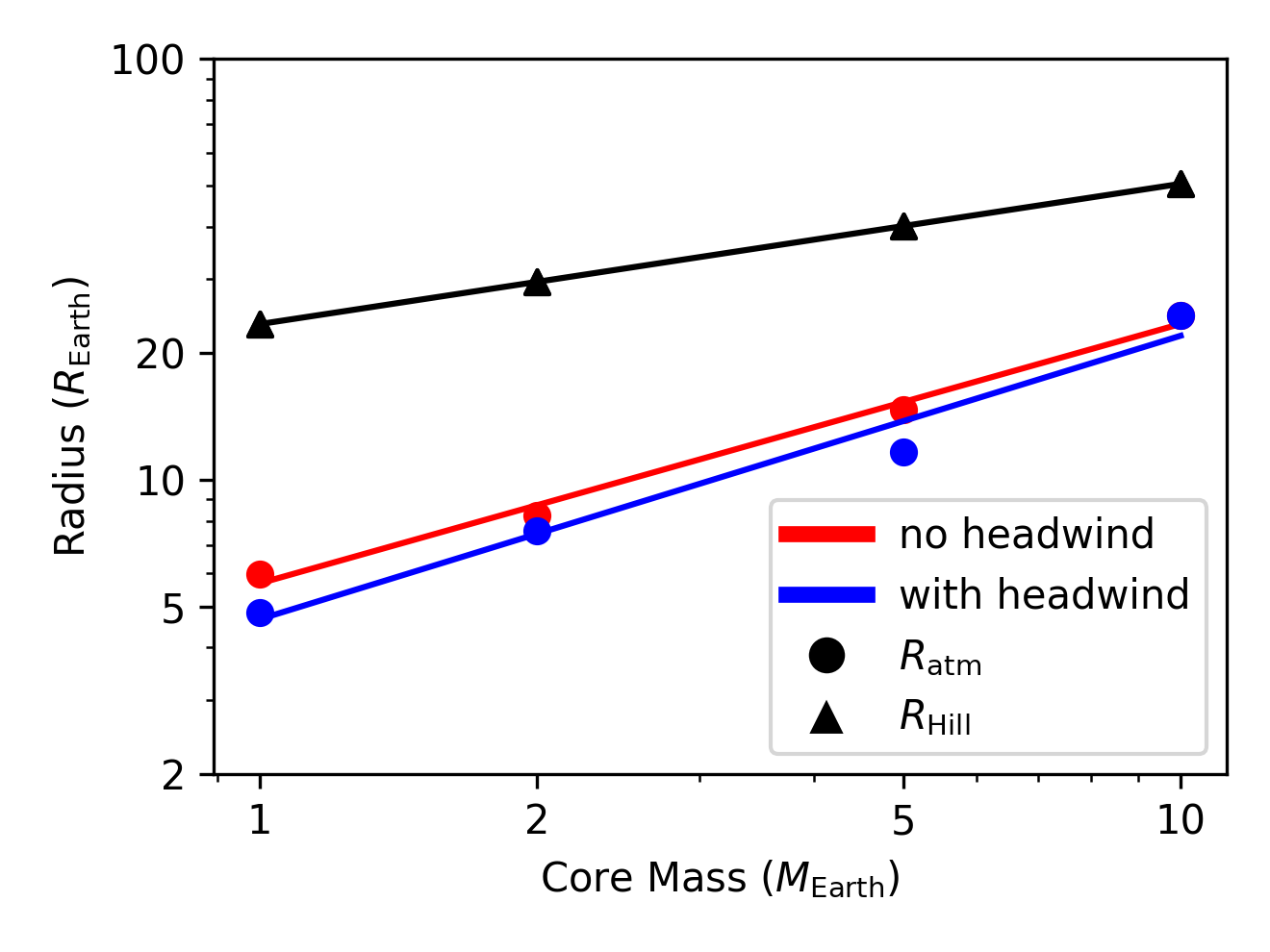}
   \caption{
      Comparison of the atmospheric radius for different core masses and headwind parameters.
      The Hill radius is by definition independent of the headwind parameter.
      We fit a power law to the data points $R_\mathrm{atm}/R_\mathrm{Earth} = a \cdot (M_c/M_\mathrm{Earth})^{n}$ using the relative least-squares method.
      When the headwind is included, we obtain $a = 4.7 \pm 0.6, n = 0.67 \pm 0.08$.
      Without the headwind, the result of the fit is $a = 5.7 \pm 0.3, n = 0.62 \pm 0.04$.
   }
   \label{fig:atmospheric_radius}%
\end{figure}

\begin{figure}[ht]
   \centering
   \includegraphics[width=0.98\hsize]{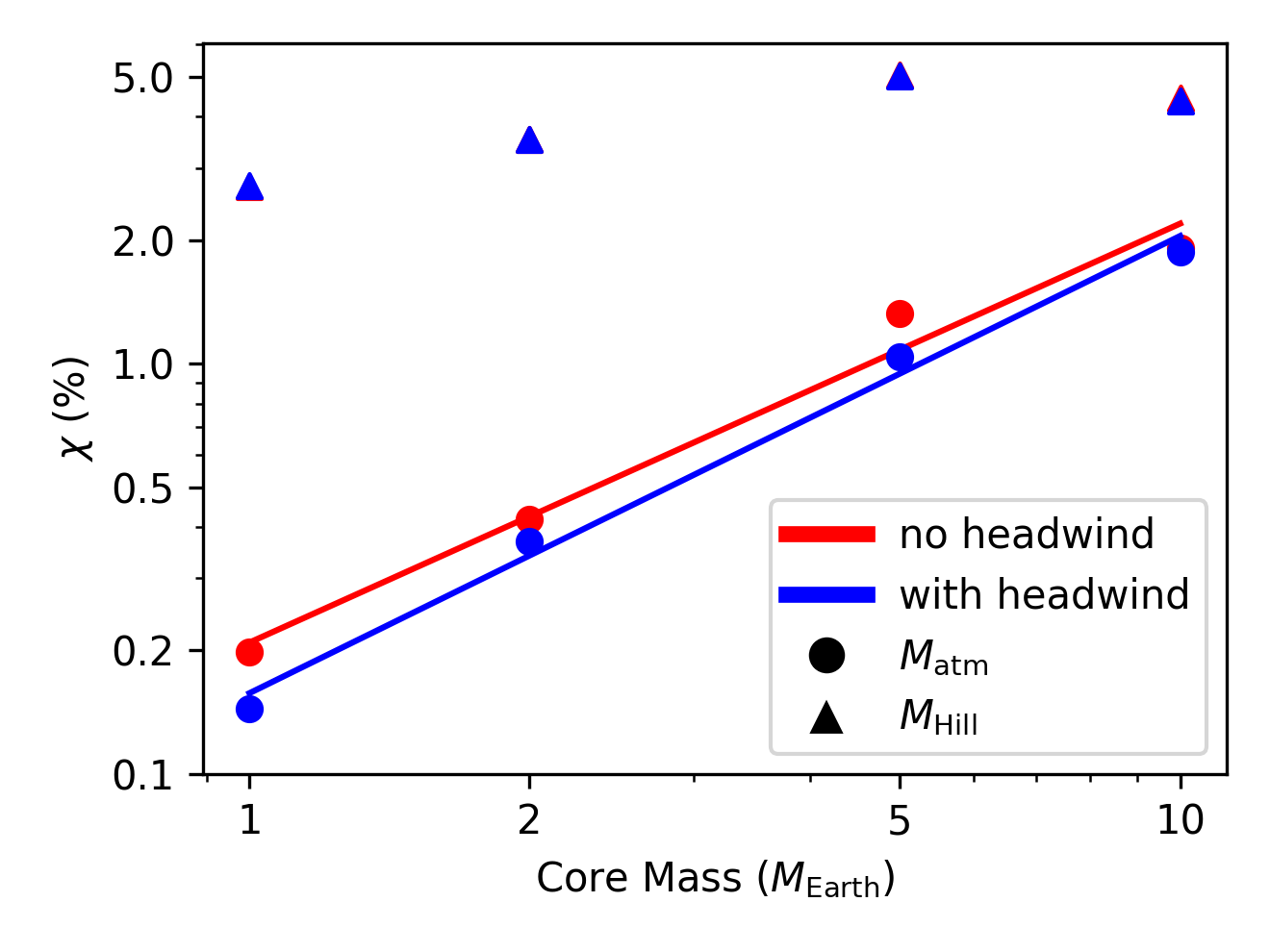}
   \caption{
      Comparison of the atmosphere-to-core mass ratio, $\chi := M_\mathrm{atm} / M_\mathrm{c}$, for different core masses and headwind parameters.
      For the mass inside the Hill sphere, the data points overlap for the different headwind parameters.
      We fit a power law to the data points $\chi = a \cdot (M_c/M_\mathrm{Earth})^{n} \, \%$ using the relative least-squares method.
      When the headwind is included we obtain $a = 0.16 \pm 0.02, n = 1.1 \pm 0.1$.
      Without the headwind, the result of the fit is $a = 0.21 \pm 0.03, n = 1.0 \pm 0.1$.
   }
   \label{fig:atmospheric_mass}%
\end{figure}

\subsection{Atmosphere size and mass}
\label{sec:atmospheric_mass}

To determine the atmospheric mass, we used the atmospheric radius $R_\mathrm{atm}$ , which we defined based on dynamical arguments measured in thermodynamic equilibrium, as explained in Sect. \ref{sec:length_scales}.
Additionally, for easy comparison with other work, we calculated the mass inside the Hill sphere and Bondi sphere.
The minimum of the Bondi and Hill radius is often quoted as an upper limit for the size of the atmosphere as the gas outside either escapes because of the stellar gravity or the thermal pressure.
However, because we did not observe a traditional gravitationally bound atmosphere, they have little physical meaning in our case besides serving as a length scale.
When we talk about the protoplanetary atmosphere, we refer to the region inside $R_\mathrm{atm}$.

Table \ref{table:simulations} shows the radii and mass of these three spheres for the different simulations.
All simulations result in atmospheric masses well below the core mass.
Hence, the envelope does not become self-gravitating and is therefore well below the critical core mass for runaway gas accretion \citep{Mizuno_1978}.
The Bondi radius quickly increases with the core mass, and the Hill radius becomes the limiting parameter for the size of the atmosphere for planetary core masses $M_c \ge 2\, M_\mathrm{Earth}$.
The size of the atmosphere measured with the described method, $R_\mathrm{atm}$, also significantly increases with core mass.
Together with the higher gas densities at higher core masses, this results in an increased atmospheric mass, $M_\mathrm{atm}$, and in an increased atmospheric mass ratio, $\chi := M_\mathrm{atm} / M_\mathrm{c}$.
The atmospheric mass and the atmospheric radius values from Table \ref{table:simulations} are visualized graphically in Figs. 14 and 15, and a power-law fit to the atmospheric radius and mass is applied.
The calculated exponents are within errors: they are the same for the headwind case and for the case without headwind.
The size of the measured atmosphere approximately scales with $M_c^{2/3}$ and the mass ratio of the atmosphere scales linearly with the core mass ($\chi \sim M_c \Rightarrow M_\mathrm{atm} \sim M_c^2$).
This means that the steeper density profile for higher core masses has little effect on the total atmospheric mass, that is, most of the atmospheric mass is contributed by the outer layers of the atmosphere.
The higher atmospheric mass at higher core masses comes from the increase in atmospheric size instead of the increase in density.

If a headwind is included, the Bondi radius becomes slightly smaller for the simulations with $M_c \ge 5$, indicating a slightly hotter atmosphere because of the increase in recycling efficiency, hence, a stronger radiative cooling and contraction as well.
We suspect that the same effect also occurs for lower core masses, but is too small to be measurable.
The Hill radius is independent of the headwind parameter by definition.
All simulations with a headwind show a smaller $R_\mathrm{atm}$ because the horseshoe orbits penetrate more deeply.
This effect significantly decreases the mass that rotates around the planetary core, $M_\mathrm{atm}$, as the atmospheric size is the dominant factor for the atmospheric mass.
At $M_c = 10 \, M_\mathrm{Earth}$ , the thermal mass is so high that this effect becomes less significant because the gravity of the planetary core starts to dominate the flow structure.
The shift of the horseshoe orbits is less noticeable compared to the size of the atmosphere at higher planetary core masses.

For the simulations with different opacities M1-HO, M1, and M1-LO, the change in the gas density is relatively small, as can be seen from the mass inside the Hill sphere, $M_\mathrm{Hill}$.
A decreased opacity, that is, a higher cooling efficiency, results in a slightly higher atmospheric mass because of the increased density.
Considering that we varied the opacity by orders of magnitude here, the change in atmospheric mass is comparatively minor.
However, if we take the change in the size of the atmosphere $R_\mathrm{atm}$ caused by the change in opacity into account, the mass of the atmosphere changes significantly.

\section{Discussion}

\subsection{Atmosphere-disk recycling and thermodynamic equilibrium}

In \cite{Moldenhauer_2021}, we showed that atmosphere-disk recycling is capable of fully compensating for radiative cooling.
With our parameter study, we here confirmed that this holds true even for higher core masses and different opacities.
At a separation of $a_p = 0.1 \, \mathrm{au,}$ the whole atmosphere is recycled.
In addition, we found that for higher core masses, the recycling pattern changes.
Instead of a full steady state with a velocity field that is constant in time, we find that for higher core masses, the atmosphere becomes turbulent.
However, from a thermodynamic point of view, the simulations still reach equilibrium, where hydrodynamic recycling fully compensates for radiative cooling.
Compared to \cite{Cimerman_2017}, the protoplanetary atmospheres in our simulations are less isolated.
Although \cite{Cimerman_2017} also observed a fully recycling atmosphere, they found a significantly sharper entropy gradient between the inner parts of the atmosphere and the disk.
We suspect that this is because a bug in their boundary conditions forced them to use gravitational smoothing and because their atmosphere is still contracting.
In this regard, our results are more similar to the 3D isothermal simulations of \cite{Bethune_2019b}, where the transition is smoother.
\cite{Kurokawa_2018} observed in their nonisothermal simulations that recycling was suppressed by a buoyancy barrier caused by the entropy gradient in the atmosphere.
In our simulations, we did not observe a buoyancy barrier that inhibits recycling of the inner regions.

\subsection{In-situ formation at 0.1 au}

For our simulations we deliberately chose a very low opacity to keep the Kelvin-Helmholtz cooling timescale short, favoring the formation of larger and more massive atmospheres.
According to our findings in section \ref{sec:result_opacity}, a higher, more realistic opacity would result in a significantly smaller and less dense atmosphere.
Thus, the atmospheric mass would be even lower.
 
At the chosen separation of $a_p = 0.1 \, \mathrm{au,}$ the recycling mechanism is so efficient that even for our highest core mass of $M_c = 10 \, M_\mathrm{Earth}$, the atmospheric mass ratio only reaches $\chi = 2 \, \%$.
Extrapolating figure \ref{fig:atmospheric_mass} suggest that for even higher core masses than the range we explored, the atmosphere-to-core mass ratio would still not grow high enough for runaway gas accretion to be possible.
This is in agreement with \cite{Fung_2019}, who found in their 3D simulations that the atmospheric mass ratio stays below $10 \, \%$ even for $M_c = 20 \, M_\mathrm{Earth}$ and an isothermal equation of state.
We suspect that the atmospheric mass ratio will increase for larger separations as the recycling timescale scales with the dynamical timescale, $\Omega_K^{-1}$.
A longer recycling timescale corresponds to a lower rate at which high-entropy circumstellar gas enters the atmosphere. Hence, the atmosphere cools more efficiently.
As the separation and thus the dynamical timescale is increased, we suspect that there will be a point at which recycling becomes too inefficient to counteract radiative cooling and more massive atmospheres become possible.
Our results show that protoplanetary cores, $M_c \le 10 M_\mathrm{Earth}$ at 0.1 au, fail to accrete atmospheres with $M_\mathrm{atm} > {\sim}2 \, \%$ when atmospheric recycling is taken into account even when the luminosity through the accretion of solids is ignored and an artificially low gas opacity is assumed.

\subsection{Rotational profile}

Because of the conservation of angular momentum, the rotational support in 2D can be calculated analytically: Gas that is accreted from the circumstellar disk at the Hill sphere has an angular momentum of $R_\mathrm{Hill}^2 \Omega_K$.
By setting this value equal to the angular momentum of the atmosphere $\sqrt{G M_p r}$, we obtain that the atmosphere is rotationally supported up to $R_\mathrm{rot} \sim R_\mathrm{Hill} / 3$ \citep{Quillen_1998}.
\cite{Ormel_2015} observed in their 2D simulations that because of vortensity conservation, the azimuthal velocity of the atmosphere increases if the gravitational smoothing length is decreased, allowing more gas to flow into the envelope.
For their smallest smoothing length, their atmosphere rotates almost at the circumplanetary Keplerian velocity.
Vortensity is not a conserved quantity in 3D.
Although we used no gravitational smoothing, all our atmosphere rotate at sub-Keplerian velocities.
The reason is the inflow of gas from the third, vertical direction.
The rotational profile of the 3D simulations in \cite{Ormel_2015b} is significantly slower than in the 2D simulations.
However, it is still dependent on the gravitational smoothing length used.
For their shortest gravitational smoothing length, their rotational profile at radii larger than the smoothing length is similar to ours.

\section{Conclusions and summary}

We conducted 3D hydrodynamic simulations with radiative transfer of a protoplanet embedded in a circumstellar disk using a local shearing box approach.
Using tracer particles and tracer fluids as post-processing, we compared the flow pattern as well as the atmospheric mass for different planetary core masses, $M_c \in [1,2,5,10] \, M_\mathrm{Earth}$.
We introduced a new radius, $R_\mathrm{atm}$, motivated by an increase in recycling timescale.
Inside $R_\mathrm{atm}$ , the recycling time of the gas is two to three orders of magnitude higher than in the rest of the Hill sphere.
Furthermore, we investigated the effect of the headwind and opacity.

Our key findings are summarized below:
For all explored parameters, the whole protoplanetary atmosphere is recycled. 
There is no (inner) region where the gas is never exchanged with the circumstellar disk.

At core masses $M_c > 2 \, M_\mathrm{Earth}$ , the atmosphere starts to become increasingly turbulent which further enhances recycling.
The size of the atmosphere, $R_\mathrm{atm}$ increases approximately proportional to $M_c^{2/3}$, that is, stronger than the Hill radius, respectively. 

The ratio of the atmosphere to the core mass, $\chi := M_\mathrm{atm}/M_c$, scales approximately linearly with the protoplanetary core mass.
Together with scaling of the atmospheric radius, $R_\mathrm{atm} \propto M_c^{2/3}$ , this means that the increase in density has only a minor effect on the atmospheric mass, that is, most mass comes from the outer layers.

The headwind parameter has only a minor effect on the size and mass of the atmosphere.
When the headwind is included, the atmosphere becomes asymmetric, slightly smaller, and therefore less massive.

While our results show that the opacity has a significant effect on the final size and mass of the protoplanetary atmosphere, we also found that the recycling timescale is not affected by the opacity.
The recycling and Kelvin-Helmholtz cooling timescale are independent of each other.
The cooling timescale is dominated by local effects such as the opacity.
We suspect that the recycling timescale will become longer at larger separations because the dynamical timescale is longer.
At the explored separation of $a_p = 0.1 \, \mathrm{au}$ even for our highest core mass of $10 \, M_\mathrm{Earth}$ , the simulation enters thermodynamic equilibrium well before an atmosphere that would support runaway gas accretion is accreted.

To conclude, atmosphere-disk recycling is an important effect capable of counterbalancing radiative cooling within a broad range of environmental conditions.
At a separation of $a_p = 0.1 \, \mathrm{au,}$ it keeps the ratio of atmosphere to core mass low and efficiently prevents the atmosphere from transitioning into the runaway gas accretion phase.

\begin{acknowledgements}
We would like to remember our co-author, colleague and friend Willy Kley who sadly passed away shortly before the acceptance of this paper.
Willy was always available for valuable scientific input and advise, and was very eager to maintain a friendly and harmonious atmosphere within his research group and with his colleagues.
He will be missed.
TWM and RK acknowledge funding through the German Research Foundation (DFG) under grant no.~KU 2849/6 as well as support through travel grants under the SPP 1992: Exoplanet Diversity program.
RK further acknowledges financial support via the Emmy Noether and Heisenberg Research Grants funded by DFG under grant no.~KU 2849/3 and 2849/9. WK acknowledges funding by the DFG through grant KL 650/31.
We acknowledge support by the High Performance and Cloud Computing Group at the Zentrum f\"ur Datenverarbeitung of the University of T\"ubingen, the state of Baden-W\"urttemberg through bwHPC and the DFG through grant no.~INST 37/935- 1 FUGG. 
\end{acknowledgements}

\bibliography{library.bib}
\bibliographystyle{aa}
\end{document}